\documentclass[preprint]{elsarticle}

\usepackage{hyperref}
\usepackage{amsfonts,amsmath,amssymb} 
\usepackage{graphicx,graphics,color}
\biboptions{sort&compress}
\journal{Physics of the Dark Universe}
\renewcommand{\(}{\begin{equation}}
\renewcommand{\)}{\end{equation}}
\newcommand{\bea}{\begin{eqnarray}}
\newcommand{\eea}{\end{eqnarray}}

\newcommand{\beq}{\begin{equation}}
\newcommand{\eeq}{\end{equation}}
\renewcommand{\(}{\begin{equation}}
\renewcommand{\)}{\end{equation}}

\usepackage{graphicx}
\usepackage{bm,natbib}
\usepackage{amsmath}
\usepackage[latin1]{inputenc}
\usepackage[english]{babel}
\usepackage[T1]{fontenc}
\usepackage{amssymb}
\usepackage{amsfonts}
\usepackage{epsfig}
\usepackage{colordvi}
\usepackage{psfrag}
\usepackage{color}

\usepackage{dcolumn}
\usepackage{multirow}
\usepackage{hyperref}
\usepackage{epstopdf}
\usepackage{bm}
\usepackage{times}
\hypersetup{
  colorlinks=true,        
  linkcolor=blue,         
  citecolor=magenta,      
}

\newcommand{\bdm}{\begin{displaymath}}
\newcommand{\edm}{\end{displaymath}}

\newcommand{\red}[1]{\textcolor{red}{#1}}
\newcommand{\be}{\begin{equation}}
\newcommand{\ee}{\end{equation}}
\newcommand{\ba}{\begin{eqnarray}}
\newcommand{\ea}{\end{eqnarray}}
\newcommand{\bc}{\begin{center}}
\newcommand{\ec}{\end{center}}
\newcommand{\epi}{$E_{\rm p,i}$}
\newcommand{\eiso}{$E_{\rm iso}$}
\newcommand{\epeiso}{$E_{\rm p,i}$--$E_{\rm iso}$}

\def\de#1/de#2{\frac{\partial {#1}}{\partial {#2}}}

\begin{document}
\bibliographystyle{article}
\begin{frontmatter}
\title{Interacting quintessence cosmology from Noether symmetries: comparing theoretical predictions with observational data}
\vspace{0.4cm}
\author[ad1,ad2]{Ester~Piedipalumbo\corref{corauth}}
\cortext[corauth]{Corresponding author}
\ead{ester@na.infn.it}
\address[ad1]{Dipartimento di Fisica "E. Pancini", Universit\`{a} degli Studi di Napoli "Federico II", 
Compl. Univ. Monte S. Angelo, Edificio 6, Via Cinthia, I-80126 Napoli, Italy}
\address[ad2]{Istituto Nazionale di Fisica Nazionale, Sez. di Napoli, Compl. Univ. Monte S. Angelo, Edificio 6, 
via Cinthia, I-80126,  Napoli, Italy}

\author[ad3]{Stefano~Vignolo}
\ead{stefano.vignolo@unige.it}
\address[ad3]{DIME, Universit\`{a} di Genova, Via all' Opera Pia 15, I-16145,  Genova, Italy}
\author[ad3]{Pasquale~Feola}
\author[ad1,ad2,ad4]{Salvatore~Capozziello}
\ead{capozziello@na.infn.it}
\address[ad4]{Scuola Superiore Meridionale, Largo S. Marcellino 10, I-80138, Napoli, Italy.}
\begin{abstract}
In the framework of scalar-tensor gravity, we consider
 non--flat interacting quintessence cosmology  where a  scalar field is interacting with dark matter. Such a scalar field can be a standard  or  a \textit{phantom} one.  We use the Noether Symmetry Approach to obtain general exact solutions for  cosmological equations and to select scalar-field  self-interaction potentials. It turns out that  the found solutions  can reproduce the accelerated expansion of  the Universe, and  are compatible with  observational dataset, as the SNeIa Pantheon data,  gamma ray bursts Hubble diagram, and  direct measurements of the Hubble parameter.
\end{abstract}
\end{frontmatter}
\section{Introduction}
\label{sec:Introduction}

The detection of the accelerated expansion of the Universe is one of the most challenging discoveries in cosmology over the last decades. To explain this  unexpected dynamics, two main proposals  have been developed. According to the first approach, the accelerated expansion is driven by some some unknown \textit{dark energy} fluid;  the second approach, instead,  is connected to non-homogeneous matter distributions or to some modification or extension of  General Relativity. According to these perspectives, several cosmological models have been proposed in literature, including a non-zero cosmological constant, standard or \textit{phantom} scalar fields, and extended/alternative theories of gravity \cite{nesseris_fr, Cai, mareknmc1,mareknmc2,nesseris, sergey,report,vasilis,ftester}.

Recently interacting dark matter - dark energy  models, dubbed as {\it interacting dark energy} or {\it coupled-dark energy}, have been proposed in different contexts in view to address several cosmological problems, such as the cosmic coincidence problem - i.e. the circumstance that   dark energy and dark matter amounts are today of the same order of magnitude, even if they evolve independently- and  cosmological tensions (~\cite{dmde,amendola,cham,chamcosmo,wei,peebles04,neil,das,das15,Bonometto17, Bonometto19,marek1, mareknmc1,sanyal, sante_interaction2018, rocco2, pdu20}.)
However, these interacting dark energy models are characterized by some phenomenological choices for the  interaction form, and there is a certain freedom in choosing  specific interaction models.
In~\citep{pdu20} we investigated, in a flat model, whether this coupling can be selected by the existence of a Noether symmetry. It turned out that this method allows us to select both the analytical form of the interaction and the self-interacting potential of the scalar field: we actually found out that the interaction term can be factorized as $F(a,\phi)= F_1(a) F_2(\phi)$. 
Moreover, we were able to obtain exact solutions of the Friedman equations, which are quite well compatible with this SNeIa data set. 
In this paper,  we extend the  {\it Noether Symmetry Approach} (see \cite{libro} for details)  to a cosmological model with non--flat spatial geometry, and to  {\it phantom} scalar fields. 

In Section \ref{Sec2}, we investigate the existence of Noether symmetries for the point-like  Lagrangian describing a single standard or  {\it phantom} scalar field coupled to  dark matter. We show that the existence of this symmetry allows a coupled dark energy field and selects the self-interaction potential leading the dark matter-dark energy interaction: we actually obtain more general expressions, and not always factorizable,  for the interaction term. 

Section \ref{Sec3} is devoted to obtain  general exact solutions  for the Friedman equations, which naturally supply accelerated expansions.

In Section \ref{Sec4} we finally  compare the theoretical solutions  with different datasets in order to achieve a reliable cosmic history at different redshifts.  
In Section \ref{conclusion} we draw conclusions. 

\section{Interacting scalar-tensor cosmology}
\label{Sec2}
The Noether Symmetry Approach \cite{libro} provides a geometric selection rule to find out the unknown parameters or functions in the gravitational action, and to solve the cosmological equations
\cite{PHRVA,RNCIB,defelice, mareknmc2, felixGB,andronikos, phantom, sanyal_ester1,sanyal_ester2,report}. Moreover, the existence of a
Noether symmetry allows to reduce the dynamical system that, in most cases, results integrable. In the present case,  let us consider the following action functional
\begin{equation}\label{2.1}
{\cal A} = \int{\sqrt{-g}\left[-\frac{1}{2}R + \frac{\epsilon}{2}g^{ij}\phi_i\phi_j + V(\phi) + \tilde{\cal L}(g_{ij},\phi)\right]d^4x}\,,
\end{equation}
describing a theory of gravity with a minimally coupled scalar field  interacting with the dark matter component. We are actually interested in investigating  cosmological models in which the dark energy component, represented by a quintessential scalar field, is directly coupled to a perfect fluid,-whose contribution has been incorporated into the Lagrangian, by using the conservations laws. That kind of perfect fluid is described by a standard term  $Ma^{-3\left(\gamma-1\right)}$ . In Eq. \eqref{2.1}, $V(\phi)$ denotes the self--interaction potential of the scalar field $\phi$, whereas 
\begin{equation}\label{2.1bis}
\tilde{\cal L}(g_{ij},\phi) = {\cal L}_m(g_{ij}) + {\cal L}_m^{int}(g_{ij},\phi)\,,
\end{equation}
is the sum of  standard matter Lagrangian function ${\cal L}_m(g_{ij})$ with  interaction term ${\cal L}_m^{int}(g_{ij},\phi)$. In order to study cosmological models deriving from the action \eqref{2.1}, let us take into account a Friedman--Robertson--Walker spacetime, whose line element is expressed as
\begin{equation}\label{2.2}
ds^2 = - dt^2 + a(t)^2\left[\frac{dr^2}{1-kr^2} + r^2d\theta^2 + r^2\sin^2\theta\/d\varphi^2\right]\,,
\end{equation}
with $k={-1,0,1}$. Inserting the content of Eq. \eqref{2.2} into Eq. \eqref{2.1}, we get the corresponding point--like Lagrangian
\begin{equation}\label{2.3}
{\cal L}(a,\phi,\dot a,\dot\phi) = 3a{\dot a}^2 -a^3\left(\frac{\epsilon{\dot\phi}^2}{2} - V(\phi)\right) - 3ka + Ma^{-3\left(\gamma-1\right)}\left(1+F(a,\phi)\right)\,,
\end{equation}
where $\gamma\in [1,2]$, and the term 
\begin{equation}
{\cal L}_m(g_{ij})=Ma^{-3\left(\gamma-1\right)}
\end{equation}
 indicates,  indeed the standard matter Lagrangian function with the constant $M$ related to the present matter density, and 
 \begin{equation}
{\cal L}_m^{int}(g_{ij},\phi) = M a^{-3\left(\gamma-1\right)}F(a,\phi)\,,
 \end{equation}
 denotes the interaction term. 
 The value of the constant  $\epsilon$ discriminates between standard and phantom quintessence field: actually, in the former case, it is $\epsilon=1$ while, in the latter,  it is $\epsilon =-1$.
The variation with respect to the two dynamical fields $a$ and $\phi$ gives the  Euler-Lagrange equations
\begin{subequations}\label{2.4}
\begin{equation}\label{2.4a}
2\frac{\ddot a}{a} + H^2 + \frac{k} {a^2} + \frac{\epsilon{\dot\phi}^2}{2} - V(\phi)  + \left(\gamma-1\right)Ma^{-3\gamma}\left(1+F\right) - \frac{1}{3}Ma^{\left(-3\gamma+1\right)}\frac{\partial F}{\partial a}=0\,,
\end{equation}
\begin{equation}\label{2.4b}
\epsilon\ddot\phi + 3\epsilon\/H\dot\phi + \frac{\partial V}{\partial\phi} + Ma^{-3\gamma}\frac{\partial F}{\partial\phi} = 0\,,
\end{equation}
\end{subequations}
where ${\displaystyle H=\frac{\dot a}{a}}$ is the Hubble parameter. Moreover, the Jacobi first integral provides  the relation
\begin{equation}
\label{2.5}
3H^2 - \frac{\epsilon{\dot\phi}^2}{2} + \frac{3k}{a^2} - V(\phi)  - Ma^{-3\gamma}\left(1+F\right) = 0\,,
\end{equation}
corresponding to the $(0,0)$ Einstein field equation.

 In the case of $\gamma =1$,   we can write  Eq.\eqref{2.5} in the  form:
\begin{equation}
3 H^2=\rho_m+\rho_k+\rho_{\phi}^{eff}\,,
\end{equation}
where  the effective energy density of the $\phi$-field is given by
\begin{equation}
\rho_{\phi}^{eff}= \rho_{\phi} +M a^{-3}F(a,\phi)\,, \label{fi-effdensity}
\end{equation}
and $\rho_{\phi}$ is the scalar field energy density
\begin{equation}\label{fi-stdensity}
\rho_{\phi}=\frac {1}{2} \epsilon\dot{\phi}^2+ V(\phi)\,.
\end{equation}
Analogously, it is possible to define an effective pressure of the scalar field as:
\begin{equation}
p_{\phi}^{eff}= \frac {1}{2}\epsilon \dot{\phi}^2- V(\phi)+ \frac{M }{3 a^2}\frac{\partial F(a,\phi)}{\partial a}\,. \label{fi-pressure}
\end{equation}
With these  two expressions, it turns out that  Eq.\eqref{2.4a} takes the form
\begin{equation}
6 \left(\frac{\ddot{a}}{a}\right)=-\left(\rho_{eff}+ 3 p_{eff}\right)\label{addot2}\,,
\end{equation}
and it is possible to define an effective equation of state 
\begin{equation}
\label{eos-eff}
w_\phi^{eff}=\displaystyle \frac{p_{\phi}}{ \rho_{\phi}}=\frac{\frac {1}{2} \epsilon\dot{\phi}^2
- V(\phi)+ \frac{M }{3 a^2}\frac{\partial F(a,\phi)}{\partial a}}{\frac {1}{2} \epsilon \dot{\phi}^2+ V(\phi)+M a^{-3}F(a,\phi)}\,,
\end{equation}
which drives the dynamics of the model.

\section{Noether symmetries  and exact solutions}
\label{Sec3}
The system of differential Eqs. \eqref{2.4a},\eqref{2.4b}, \eqref{2.5} is  non-linear and many choices are possible for the interaction term $ F(a,\phi)$ and  the self interaction potential $V(\phi)$. In order to solve the system, we search for Noether symmetries for the Lagrangian in Eq. \eqref{2.3}, by which it is possible to simplify the study of the dynamics. Furthermore, the existence of these symmetries allows us to fix  the forms of $ F(a,\phi)$ and $V(\phi)$.
\subsection{The case of a standard scalar field}
The configuration space  of the model is given by the  local coordinates. It is ${\cal Q}\equiv\{a,\phi\}$. The associated tangent bundle is then given by the fibered coordinates, i.e. $T{\cal Q}\equiv\{a,\phi,\dot a,\dot\phi\}$. The resulting point--like Lagrangian is a function on $T{\cal Q}$ having the local expression
\begin{equation}\label{Lagrangian}
{\cal L}(a,\phi,\dot a,\dot\phi) = 3a{\dot a}^2 -a^3\left(\frac{{\dot\phi}^2}{2} - V(\phi)\right) - 3ka + Ma^{-3\left(\gamma-1\right)}\left(1+F(a,\phi)\right)\,.
\end{equation}
We look for Noether symmetries of the Lagrangian \eqref{Lagrangian} of the form
\begin{equation}\label{Noether_symmetry}
X= \alpha\,\frac{\partial}{\partial a} + \beta\,\frac{\partial}{\partial\phi} + \dot\alpha\,\frac{\partial}{\partial{\dot a}} + \dot\beta\,\frac{\partial}{\partial{\dot\phi}}
\end{equation}
where $\alpha=\alpha(a,\phi)$, $\beta=\beta(a,\phi)$, $\dot\alpha=\dot{a}\de{\alpha}/de{a}+\dot{\phi}\de{\alpha}/de{\phi}$ and $\dot\beta=\dot{a}\de{\beta}/de{a}+\dot{\phi}\de{\beta}/de{\phi}$. The vector fields \eqref{Noether_symmetry} are defined on the tangent bundle $T{\cal Q}$ and are the canonical lift of corresponding vector fields 
\begin{equation}\label{Base_vector}
Y=\alpha\,\frac{\partial}{\partial a} + \beta\,\frac{\partial}{\partial\phi}
\end{equation}
defined on ${\cal Q}$. A Noether symmetry \eqref{Noether_symmetry} is achieved if the condition

\be\label{requirement_NS}
L_{X}{\cal L}=X{\cal L}=\alpha\de{\cal L}/de{a} + \beta\de{\cal L}/de{\phi} + \dot\alpha\de{\cal L}/de{\dot a} + \dot\beta\de{\cal L}/de{\dot\phi}=0\,,
\ee
holds, where $L_{X}$ is the Lie derivative.  The Noether charge $\Sigma_{0}$ can be written as $i_{X}\Theta_{\cal L}=\Sigma_{0}$, where $i_{X}\Theta_{\cal L}$ is the Cartan one-form.

Condition \eqref{requirement_NS} gives rise to the following system of first--order partial differential equations
\begin{subequations}\label{Noehter_equations}
\begin{equation}\label{Noether_equations_1}
\alpha + 2a\frac{\partial\alpha}{\partial a} =0\,,
\end{equation}
\begin{equation}\label{Noether_equations_2}
3\alpha + 2a\frac{\partial\beta}{\partial\phi} =0\,,
\end{equation}
\begin{equation}\label{Noether_equations_3}
6\frac{\partial\alpha}{\partial\phi} - a^2\frac{\partial\beta}{\partial a} =0\,,
\end{equation}
\begin{eqnarray}\label{Noether_equations_4}
&& 3\alpha\/a^2V(\phi) + \beta\/a^3\frac{dV}{d\phi}(\phi) -3(\gamma-1)\/M\/a^{-3\gamma+2}\alpha\left(1+F(a,\phi)\right) \nonumber \\
&&+M\/a^{-3(\gamma-1)}\frac{\partial F}{\partial a}\alpha + M\/a^{-3(\gamma-1)}\frac{\partial F}{\partial\phi}\beta - 3k\alpha=0\,.
\end{eqnarray}
\end{subequations}
 Actually the contraction appearing  in  \eqref{requirement_NS} defines a quadratic form in $\dot{a}$ and $\dot{\phi}$: in order to solve, at any value of $t$, the Eq.$L_{X}{\cal L}=0$, we equate to zero the coefficients in $\dot{\phi}, \dot{a}\dot{\phi}$, obtaining the Eqs.  \eqref{Noether_equations_1} - \eqref{Noether_equations_4}. 
It turns out that we can immediately find a not factorized particular solutions of Eqs. \eqref{Noether_equations_1}, \eqref{Noether_equations_2} and \eqref{Noether_equations_3}, given by 
\begin{subequations}\label{Noehter_equ-part}
\begin{equation}\label{equ_1}
\alpha(a,\phi)=0\,,
\end{equation}
\begin{equation}\label{equ_2}
\beta(a,\phi)=\beta_0\,,
\end{equation}
\end{subequations}
where $\beta_0$ is constant. In this case the corresponding symmetry is the field
\begin{equation}\label{field-particular}
X=\beta_0 \frac{\partial}{\partial \phi}
\end{equation}
and the constant of motion is
\begin{equation}\label{charge-particular}
\Sigma_0= a^3 \dot{\phi}\,.
\end{equation}
A more general solution of the system \eqref{Noether_equations_1}, \eqref{Noether_equations_2} and \eqref{Noether_equations_3}, can be obtained by applying the separating variables method. Actually these solutions are
\begin{subequations}\label{solutions_alpha_beta}
\begin{equation}\label{solution_alpha}
\alpha = \frac{A\/e^{\frac{1}{2}\sqrt{3/2}\phi} + B\/e^{-\frac{1}{2}\sqrt{3/2}\phi}}{a^{1/2}}
\end{equation}
\begin{equation}\label{solution_beta}
\beta= \frac{-A\sqrt{6}\/e^{\frac{1}{2}\sqrt{3/2}\phi} + B\sqrt{6}\/e^{-\frac{1}{2}\sqrt{3/2}\phi}}{a^{3/2}}
\end{equation}
\end{subequations}
where $A$ and $B$ are intergration constants.   
After, we can split Eqs. \eqref{Noether_equations_4} into two separate equations for $V(\phi)$ and $F(a,\phi)$ respectively:
\begin{subequations}\label{equation_split}
\begin{equation}\label{equation_split_1}
3\alpha\/a^2V(\phi) + \beta\/a^3V'(\phi) =0
\end{equation}
and
\begin{equation}\label{equation_split_2}
-3(\gamma-1)\/M\/a^{-3\gamma+2}\alpha\left(1+F(a,\phi)\right) + M\/a^{-3(\gamma-1)}\frac{\partial F}{\partial a}\alpha + M\/a^{-3(\gamma-1)}\frac{\partial F}{\partial\phi}\beta - 3k\alpha=0\,.
\end{equation}
\end{subequations}
Therefore, making use of Eqs. \eqref{solutions_alpha_beta}, we find  the potential
 \begin{equation}
\label{solution_Vcomplete}
V(\phi)= e^{-\sqrt{\frac{3}{2}} \phi } V_0 \left(B-A e^{\sqrt{\frac{3}{2}} \phi }\right)^2\,.
\end{equation}
Some comments are in order here.  These kind of potential is physically relevant because it gives accelerated expansion also for the inflationary paradigm \cite{power-law}. Moreover it turns out that in presence of exponential potentials it is possible to exhibit alternative Lagrangians for the Einstein field equations\citep{exp-ester}.
Furthermore, it is possible to show that exponential forms for the potential result in invertible  conformal transformations and can be related to $f(R)$ gravity in the Einstein frame \cite{curvature}.
In the following we set  $A=0$, so that the potential takes the form:
\begin{subequations}\label{solutions_V_F}
\begin{equation}\label{solution_V}
V(\phi)= V_0\/e^{-\sqrt{3/2}\phi}\,.
\end{equation}
The function $F(a,\phi)$ is 
\begin{equation}\label{solution_F}
F(a,\phi)=a^{3(\gamma-1)}G(-\sqrt{6}\ln\/a + \phi) + \frac{3k}{M}a^{3\gamma-2} - 1\,,
\end{equation}
\end{subequations}
where $G(x)$ is an arbitrary function of its own argument (therefore in the case of Eq. \eqref{solution_F} $x=-\sqrt{6}\ln\/a + \phi$).  Inserting Eqs. \eqref{solution_V} and \eqref{solution_F} into Eq. \eqref{Lagrangian}, Lagrangian \eqref{Lagrangian} assumes the expression
\begin{equation}\label{new_Lagrangiangen}
{\cal L}=a^3 \left(V_0 e^{-\sqrt{\frac{3}{2}} \phi }-\frac{\dot{\phi }^2}{2}\right)+M
   G\left(\phi -\sqrt{6} \log (a)\right)+3 \dot{a}^2 a\,.
\end{equation}
 It is worth noticing that the above symmetries  give rise to an interaction term which actually cancels the contributions due to the spatial curvature and reduces the dynamic effect of the cosmological fluid to that of a dust. The effects of the curvature remain indeed in the evolution of the scalar field, as it can be inferred from the definition of the effective density and pressure of the scalar field (see Eqs. \eqref{fi-effdensity} and \eqref{fi-pressure}).  
Moreover it turns out  that, in the case of an arbitrary form for the function  $G\left(\phi -\sqrt{6} \log (a)\right)$ in Eq. \eqref{new_Lagrangiangen}, the interaction function $F(a,\phi)$ is not  separable on $a$ and $\phi$: just for some special choices of this arbitrary term, we find out that $F(a,\phi)$ can be considered factorized.
For instance, by choosing $G(x)=Qe^{-\frac{hx}{\sqrt{6}}}$ with $Q$ and $h$ appropriate constants, we have the function 
\begin{equation}\label{particular_F}
F(a,\phi)=Qa^{3(\gamma-1)+h}\/e^{-\frac{h\phi}{\sqrt{6}}} + \frac{3k}{M}a^{3\gamma-2} - 1\,.
\end{equation}
In this case, the Lagrangian \eqref{Lagrangian} assumes the physically relevant expression
\begin{equation}\label{new_Lagrangian}
{\cal L}= 3a{\dot a}^2 -a^3\left(\frac{{\dot\phi}^2}{2} - V_0\/e^{-\sqrt{3/2}\phi}\right) + MQa^h\/e^{-\frac{h\phi}{\sqrt{6}}}\,.
\end{equation}
 We see that, once that we have been able to assign the functions $V(\phi)$, and $F(\phi)$,  it is always possible to transform the Lagrangian
in Eq. \eqref{new_Lagrangian}: 
\begin{equation}\label{transformedL}
{\cal L}(a,\phi,\dot{a},\dot{\phi})\rightarrow {\cal \tilde{L}}(u,\dot{u},\dot{v})\,,
\end{equation} 
i.e. v becomes a cyclic variable for the transformed Lagrangian. Actually it is well known that under a point transformation, a vector field $X$ becomes
\begin{equation}\label{transformedX}
\tilde{X}= \left(i_{X} dQ^{k} \right)\frac{\partial}{\partial Q^k}+\frac{d}{d t}\left(i_{X} dQ^{k} \right) \frac{\partial}{\partial \dot{Q^k}}\,.
\end{equation} 
If $X$ is a Noether symmetry, and we consider a point transformation in a such a way that 
\begin{eqnarray}\label{transformed_coordinates}
&&i_X dQ^1=1\,,\\
&& i_XdQ^{j}=0\, ( j \ne1)\,,
\end{eqnarray}
we obtain 
\begin{subequations}\label{transformedX2}
\begin{equation}\label{transfX1}
\tilde{X}= \frac{\partial}{\partial Q^1}\,,
\end{equation}
\begin{equation}\label{transfX2}
\frac{\partial {\cal \tilde{L}} }{\partial Q^1}=0\,.
\end{equation}
\end{subequations} 
Therefore $Q^1$ is a cyclic coordinate for $ {\cal \tilde{L}} $, and the dynamics can be simplified according to a well known procedure, which can often allow us to obtain exact solutions for the associated Euler-Lagrangian equations. It is worth remarking that the change of coordinates defined in Eq.\eqref{transfX1}-\eqref{transfX2}is not unique  (so that an appropriate choice is important), and the solution of this system  it is, in general, not defined everywhere in the space. In the case of our Lagrangian \eqref{new_Lagrangian}, we perform the change of variables:
\begin{subequations}\label{transformed_cosmocoordinate}
\begin{equation}\label{cosmocoordinate1}
i_X dv=\alpha \frac{\partial v}{\partial a}+\beta \frac{\partial v}{\partial \phi}=1\,,
\end{equation}
\begin{equation}\label{cosmocoordinate2}
i_X dv= \alpha \frac{\partial u}{\partial a}+\beta \frac{\partial u}{\partial \phi}=0\,.
\end{equation}
\end{subequations}
 It turns out that $v$ is cyclic for the transformed Lagrangian (i.e. $\displaystyle\frac{\partial{\cal L}}{\partial v}=0$). \footnote{ In the following, with abuse of notation, we shall write ${\cal L}$ and not $ {\cal \tilde{L}}$ to indicate the transformed Lagrangian.} We solve Eqs. \eqref{cosmocoordinate1}-\eqref{cosmocoordinate2}
and obtain 
\begin{subequations}\label{change_coordinates}
\begin{equation}\label{change_coordinates_1}
a=\left(uv\right)^{1/3}\,,
\end{equation}
\begin{equation}\label{change_coordinates_2}
\phi=-\sqrt{2/3}\ln{\frac{u}{v}}\,.
\end{equation}
\end{subequations}
In terms of the new coordinates,  the  Lagrangian \eqref{new_Lagrangian} can be expressed as
\begin{equation}\label{new_Lagrangian_bis}
{\cal L}=\frac{4}{3}\dot{u}\dot{v} + V_0\/u^2 + MQu^{\frac{2h}{3}}\,,
\end{equation}
where $v$ is a cyclic coordinate.
The associated conserved momentum is given by
\begin{equation}\label{conserved_current}
\frac{\partial{\cal L}}{\partial{\dot v}}=\frac{4}{3}\dot u =\Sigma\,,
\end{equation}
where $\Sigma$ is a constant of motion.
By integrating Eq.\eqref{conserved_current}, we get
\begin{equation}\label{solution_u}
u(t)=\frac{3}{4}\Sigma\/t + u_0\,.
\end{equation}
Using  the Jacobi first integral of \eqref{new_Lagrangian_bis}, we obtain the evolution equation for $v$
\begin{equation}\label{equation_v}
\dot v= \frac{V_0}{\Sigma}\left(\frac{3}{4}\Sigma\/t + u_0\right)^2 + \frac{MQ}{\Sigma}\left(\frac{3}{4}\Sigma\/t + u_0\right)^{\frac{2h}{3}}
\end{equation}
which is directly integrated, giving rise to
\begin{equation}\label{solution_v}
v(t)=\frac{4V_0}{9\Sigma^2}\left(\frac{3}{4}\Sigma\/t + u_0\right)^3 + \frac{4MQ}{3\Sigma^2\left(\frac{2h}{3}+1\right)}\left(\frac{3}{4}\Sigma\/t + u_0\right)^{\frac{2h}{3}+1} + C\,.
\end{equation}
Therefore, Eqs. \eqref{change_coordinates_1} and \eqref{change_coordinates_2} provide the analytical form for the scale factor, $a(t)$, and the scalar field, $\phi(t)$ in terms of  $u(t)$ and $v(t)$.

Another possible choice  in Eq. \eqref{solution_F} is  $G(x)=Q\/x^n$. The point--like Lagrangian  \eqref{Lagrangian} becomes 
\begin{equation}\label{Lagrangian_bis}
{\cal L}= 3a{\dot a}^2 -a^3\left(\frac{{\dot\phi}^2}{2} - V_0\/e^{-\sqrt{3/2}\phi}\right) + MQ\/(-\sqrt{6}\ln\/a + \phi)^n\,.
\end{equation}
By performing again the coordinate trasformation \eqref{change_coordinates}, the Lagrangian \eqref{Lagrangian_bis} assumes the form
\begin{equation}\label{Lagrangian_bis_new}
{\cal L}=\frac{4}{3}{\dot u}{\dot v} + V_0\/u^2 + 6^{n/2}MQ\left(-\frac{2}{3}\ln\/u\right)^n
\end{equation}
Again, the variable $v$ is cyclic, the quantity ${\displaystyle \frac{\partial{\cal L}}{\partial \dot v}=\frac{4}{3}\dot u}$ is conserved and the solution \eqref{solution_u} holds. By inserting solution \eqref{solution_u} into the Jacobi first integral, we get the evolution equation for the variable $v$ 
\begin{equation}\label{equation_v_bis}
\dot v = \frac{V_0}{\Sigma}\left(\frac{3}{4}\Sigma\/t + u_0\right)^2 + 6^{n/2}\frac{MQ}{\Sigma}\left(-\frac{2}{3}\ln\left(\frac{3}{4}\Sigma\/t + u_0\right)\right)^n
\end{equation}
which is solved as
\begin{equation}\label{solution_v_bis}
v(t)= \frac{4V_0}{9\Sigma^2}\left(\frac{3}{4}\Sigma\/t + u_0\right)^3 + \int{6^{n/2}\frac{MQ}{\Sigma}\left(-\frac{2}{3}\ln\left(\frac{3}{4}\Sigma\/t + u_0\right)\right)^n}\,dt
\end{equation}
Similar solutions are obtained by choosing $G(x)=Q\/e^x$, $G(x)=Qln\/(x)$ or $G(x)=Q\/\sqrt{x}$ in Eq. \eqref{solution_F}. For $G(x)=Q\/e^x$, the corresponding point--like Lagrangian is given by
\begin{equation}\label{Lagrangian_tris}
{\cal L}= 3a{\dot a}^2 -a^3\left(\frac{{\dot\phi}^2}{2} - V_0\/e^{-\sqrt{3/2}\phi}\right) + MQ\/e^{-\sqrt{6}\ln\/a + \phi}\,,
\end{equation}
and the transformed  Lagrangian \eqref{Lagrangian_tris} becomes
\begin{equation}\label{Lagrangian_tris_new}
{\cal L}=\frac{4}{3}{\dot u}{\dot v} + V_0\/u^2 + MQ\/u^{-\sqrt{8/3}}\,.
\end{equation} 
Also in this case the variable $v$ is cyclic and the quantity ${\displaystyle \frac{\partial{\cal L}}{\partial \dot v}=\frac{4}{3}\dot u}$ is conserved. The solution \eqref{solution_u} is still valid and we can use it into the Jacobi first integral. The evolution equation for the variable $v$ is 
\begin{equation}\label{equation_v_tris}
\dot v = \frac{V_0}{\Sigma}\left(\frac{3}{4}\Sigma\/t + u_0\right)^2 + \frac{MQ}{\Sigma}\left(\frac{3}{4}\Sigma\/t + u_0\right)^{-\sqrt{8/3}}
\end{equation}
and it admits the solution
\begin{equation}\label{solution_v_tris}
v(t)=\frac{4V_0}{9\Sigma^2}\left(\frac{3}{4}\Sigma\/t + u_0\right)^3 + \frac{4MQ}{3\Sigma^2\left(-\sqrt{8/3} +1\right)}\left(\frac{3}{4}\Sigma\/t + u_0\right)^{-\sqrt{8/3} +1} + C
\end{equation}
In the case $G(x)=Q\/ln(x)$, the corresponding point--like Lagrangian is 
\begin{equation}\label{Lagrangian_ln}
{\cal L}= 3a{\dot a}^2 -a^3\left(\frac{{\dot\phi}^2}{2} - V_0\/e^{-\sqrt{3/2}\phi}\right) + MQ\/ln(-\sqrt{6}\ln\/a + \phi)\,,
\end{equation}
and the transformed Lagrangian \eqref{Lagrangian_ln} assumes the form:
\begin{equation}\label{Lagrangian_ln_new}
{\cal L}=\frac{4}{3}{\dot u}{\dot v} + V_0\/u^2 + + MQ\/ln(-\sqrt{8/3}\ln\/u)
\end{equation} 
Since $v$ is again a cyclic variable, the quantity ${\displaystyle \frac{\partial{\cal L}}{\partial \dot v}=\frac{4}{3}\dot u}$ is conserved and we get solution \eqref{solution_u} again. Inserting \eqref{solution_u} into the Jacobi first integral, we get the evolution equation for the variable $v$
\begin{equation}\label{equation_v_ln}
\dot v = \frac{V_0}{\Sigma}\left(\frac{3}{4}\Sigma\/t + u_0\right)^2 + \frac{MQ}{\Sigma}\/ln\left(-\sqrt{8/3}\ln\/\left(\frac{3}{4}\Sigma\/t + u_0\right)\right)
\end{equation}
which has solution
\begin{equation}\label{solution_v_ln}
v(t)= \frac{4V_0}{9\Sigma^2}\left(\frac{3}{4}\Sigma\/t + u_0\right)^3 + \int{\frac{MQ}{\Sigma}\/ln\left(-\sqrt{8/3}\ln\/\left(\frac{3}{4}\Sigma\/t + u_0\right)\right)}\,dt
\end{equation}
Finally, for $G(x)=Q\/\sqrt{x}$ the point--like Lagrangian  is
\begin{equation}\label{Lagrangian_sqrt}
{\cal L}= 3a{\dot a}^2 -a^3\left(\frac{{\dot\phi}^2}{2} - V_0\/e^{-\sqrt{3/2}\phi}\right) + MQ\sqrt{(-\sqrt{6}\ln\/a + \phi)}
\end{equation}
 the transformed Lagrangian \eqref{Lagrangian_sqrt} can be written as
\begin{equation}\label{Lagrangian_sqrt_new}
{\cal L}=\frac{4}{3}{\dot u}{\dot v} + V_0\/u^2 + + MQ\sqrt{-\sqrt{8/3}\ln\/u}\,.
\end{equation}
Due to the conservation of momentum ${\displaystyle \frac{\partial{\cal L}}{\partial \dot v}=\frac{4}{3}\dot u}$, the solution \eqref{solution_u} holds again, while the Jacobi first integral yields the evolution equation for $v$ 
\begin{equation}\label{equation_v_sqrt}
\dot v = \frac{V_0}{\Sigma}\left(\frac{3}{4}\Sigma\/t + u_0\right)^2 + \frac{MQ}{\Sigma}\sqrt{-\sqrt{8/3}\ln\/\left(\frac{3}{4}\Sigma\/t + u_0\right)}
\end{equation}
The latter can be integrated as 
\begin{eqnarray}
\label{solution_v_sqrt}
&&v(t) = \frac{4V_0}{9\Sigma^2}\left(\frac{3}{4}\Sigma\/t + u_0\right)^3 + \int{\frac{MQ}{\Sigma}\sqrt{-\sqrt{8/3}\ln\/\left(\frac{3}{4}\Sigma\/t + u_0\right)}}\,dt =\nonumber \\
&& \frac{4V_0}{9\Sigma^2}\left(\frac{3}{4}\Sigma\/t + u_0\right)^3 + \\ && \frac{2^{3/4} \sqrt{-\log \left(\frac{3}{4} \left(\Sigma  t+u_0\right)\right)}
   \left(-\frac{\left(\Sigma  t+u_0\right) {\cal F}\left(\sqrt{\log \left(\frac{3}{4} \left(t
   \Sigma +u_0\right)\right)}\right)}{\sqrt{\log \left(\frac{3}{4} \left(\Sigma 
   t+u_0\right)\right)}}+\Sigma  t+u_0\right)}{\sqrt[4]{3} \Sigma }\nonumber\,.
\end{eqnarray}
Here ${\cal F}$ is a  Dawson integral, defined as:
\begin{equation}
{\cal F}(x)= e^{-x^2} \int^x_0 e^{y^2} dy\,. 
\end{equation} 
It is worth stressing that the  Dawson integral can be represented in terms of the  imaginary error function:
\begin{equation}
{\cal F}(x)= \frac{1}{2} \sqrt{\pi } e^{-x^2} \frac{erf(ix)}{i}\,.
\end{equation}
It is worth noting that the solutions corresponding to the symmetries of the Lagrangians described in Eqs.\eqref{Lagrangian_bis}, \eqref{Lagrangian_tris}, \eqref{Lagrangian_ln} and \eqref{Lagrangian_sqrt} are expressed only in terms of the new variables $u(t)$ and $v(t)$, for reason of mathematical clarity and simplicity: the use of the old variables ($a(t)$ and $\phi(t)$) would result in very long and undesirably complicated expressions. Moreover, as final remark, we emphasize that the Noether symmetry approach allowed us to  find a considerable number of symmetries and related cosmological solutions: in order to understand if these solutions provide available candidates for describing the dark energy component, all these mathematical solutions ought to be analyzed in detail, and case by case, from a physical point of view -checking, for example, whether they give rise to accelerated expansion of the Universe. In a forthcoming paper we plat to analyze and characterized  all the mathematical solutions provided by the Noether symmetry method on the base of their physical features. Here, to illustrate our method, we concentrate our attention on the solution in Eqs. \eqref{solution_u}, and \eqref{solution_v} concerning the case of the standard scalar field, and on the solution in Eqs.  \eqref{EL-solutions_u_tris_a} and \eqref{EL-solutions_v_tris_b}, concerning the case of a phantom scalar field, as we shall discuss below. 
\subsection{The case of a phantom field }
 Almost all data sets from cosmological probes are compatible with dark energy equations of state parameter where $w<-1$ (see for instance \citep{Planck2020}): dark energy with this kind of equations of state is often called phantom dark energy. Phantom fluids were first introduced by Caldwell, who suggested the name due to the circumstance that phantoms or ghosts possess negative energy, which leads to instabilities on both classical and quantum level \citep{CaldwellPh, CarrollPh, Noijri}, and violate the energy conditions. From the theoretical point of view, however, contexts with phantom-like equations of state, which do not lead to energy conditions violation, have been explored. Actually, phantom type of matter was investigated in several cosmological scenarios \citep{Ludwick2017,Nojiri2006, Vazquez2021,Bargach2021,Kucukakca2020}. Here we consider phantom interacting dark energy, and look for general  analytical form of the interaction and the self-interacting potential of the phantom field.
The point--like Lagrangian is now of the form 
\begin{equation}\label{Lagrangianbis}
{\cal L}(a,\phi,\dot a,\dot\phi) = 3a{\dot a}^2 + a^3\left(\frac{{\dot\phi}^2}{2} + V(\phi)\right) - 3ka + Ma^{-3\left(\gamma-1\right)}\left(1+F(a,\phi)\right)
\end{equation}
In this case, the conditions for the existence of a  Noether symmetry \eqref{requirement_NS} are
\begin{subequations}\label{Noehter_equationsbis}
\begin{equation}\label{Noether_equations_1bis}
\alpha + 2a\frac{\partial\alpha}{\partial a} =0\,,
\end{equation}
\begin{equation}\label{Noether_equations_2bis}
3\alpha + 2a\frac{\partial\beta}{\partial\phi} =0
\end{equation}
\begin{equation}\label{Noether_equations_3bis}
6\frac{\partial\alpha}{\partial\phi} + a^2\frac{\partial\beta}{\partial a} =0\,,
\end{equation}
\begin{eqnarray}\label{Noether_equations_4bis}
&&3\alpha\/a^2V(\phi) + \beta\/a^3\frac{dV}{d\phi}(\phi) -3(\gamma-1)\/M\/a^{-3\gamma+2}\alpha\left(1+F(a,\phi)\right)  \\ \nonumber+ && M\/a^{-3(\gamma-1)}\frac{\partial F}{\partial a}\alpha+ M\/a^{-3(\gamma-1)}\frac{\partial F}{\partial\phi}\beta - 3k\alpha=0\,.
\end{eqnarray}
\end{subequations}
Once again by separating variables, Eqs. \eqref{Noether_equations_1bis}, \eqref{Noether_equations_2bis} and \eqref{Noether_equations_3bis}  have solutions of the form
\begin{subequations}\label{solutions_alpha_betabis}
\begin{equation}\label{solution_alphabis}
\alpha = \frac{-A\cos\left(\frac{\sqrt{6}}{4}\phi\right) + B\sin\left(\frac{\sqrt{6}}{4}\phi\right)}{a^{1/2}}\,,
\end{equation}
\begin{equation}\label{solution_betabis}
\beta= \frac{A\sqrt{6}\sin\left(\frac{\sqrt{6}}{4}\phi\right) + B\sqrt{6}\cos\left(\frac{\sqrt{6}}{4}\phi\right)}{a^{3/2}}\,,
\end{equation}
\end{subequations}
where $A$ and $B$ are appropriate integration constants. By splitting again Eq. \eqref{Noether_equations_4bis} into
\begin{subequations}\label{equation_splitbis}
\begin{equation}\label{equation_split_1bis}
3\alpha\/a^2V(\phi) + \beta\/a^3V'(\phi) =0\,,
\end{equation}
and
\begin{equation}\label{equation_split_2bis}
-3(\gamma-1)\/M\/a^{-3\gamma+2}\alpha\left(1+F(a,\phi)\right) + M\/a^{-3(\gamma-1)}\frac{\partial F}{\partial a}\alpha + M\/a^{-3(\gamma-1)}\frac{\partial F}{\partial\phi}\beta - 3k\alpha=0\,,
\end{equation}
\end{subequations}
and, using Eqs. \eqref{solutions_alpha_betabis} ,
 it turns out that 
\begin{equation}\label{Vphantom}
V(\phi)=V_0 \left(B \cos \left(\frac{1}{2} \sqrt{\frac{3}{2}} \phi \right)-A \sin
   \left(\frac{1}{2} \sqrt{\frac{3}{2}} \phi \right)\right)^2\,.
   \end{equation}
   Also in this case we can set  $A=0$ for simplicity, and  we obtain solutions for $V(\phi)$ and $F(a,\phi)$ of the form
\begin{subequations}\label{solutions_V_Fbis}
\begin{equation}\label{solution_Vbis}
V(\phi)= V_0\/\cos^2\left(\frac{\sqrt{6}}{4}\phi\right)\,,
\end{equation}
\begin{equation}\label{solution_Fbis}
F(a,\phi)=a^{3(\gamma-1)}G\left(\cos\left(\frac{\sqrt{6}}{4}\phi\right)a^{\frac{3}{2}}\right) + \frac{3k}{M}a^{3\gamma-2} - 1\,,
\end{equation}
\end{subequations}
where $G(x)$ is an arbitrary function of its own argument (therefore in the case of Eq. \eqref{solution_Fbis} $x=\cos\left(\frac{\sqrt{6}}{4}\phi\right)a^{\frac{3}{2}}$. It is worth noting that the choice  $A=0$ implies a periodic the self-interaction potential, but not a periodic interaction term, due to the presence of the term $a^{\frac{3}{2}}$. As a consequence, differently from the case of a non-interacting phantom model-when a periodic potential actually gives rise to oscillating solutions \citep{phantom,exp-ester} - whatever the analytical form of $G(x)$, we have no oscillating cosmological solutions (see, for instance, Eqs. \eqref{a},\eqref{phi},\eqref{a},\eqref{EL-solutions_u},\eqref{EL-solutions_v} below). Moreover, also in the case of a phantom scalar field, it turns out that the  interaction term  effectively cancels the contributions due to the spatial curvature and reduces the dynamic effect of the cosmological fluid to that of  dust.

 For instance, by choosing $G(x)=x$ and inserting expressions \eqref{solutions_V_Fbis} into Eq. \eqref{Lagrangianbis}, the point--like Lagrangian assumes the form
\begin{equation}\label{new_Lagrangianbis}
{\cal L}= 3a{\dot a}^2 + a^3\left(\frac{{\dot\phi}^2}{2} + V_0\/\cos^2\left(\frac{\sqrt{6}}{4}\phi\right)\right) + M\cos\left(\frac{\sqrt{6}}{4}\phi\right)a^{\frac{3}{2}}
\end{equation}
 Therefore, also in the case of  phantom scalar field,  we can look for new coordinates $u$ and $v$, solutions of Eqs.:
\begin{subequations}\label{transformed_cosmocoordinateph}
\begin{equation}\label{cosmocoordinate1ph}
i_X dv=\alpha \frac{\partial v}{\partial a}+\beta \frac{\partial v}{\partial \phi}=1\,,
\end{equation}
\begin{equation}\label{cosmocoordinate2ph}
i_X dv= \alpha \frac{\partial u}{\partial a}+\beta \frac{\partial u}{\partial \phi}=0\,.
\end{equation}
\end{subequations}

Again, under the hypothesis $A=0$, they are given by
\begin{subequations}\label{new_coordinates}
\begin{equation}\label{new_coordinates_u}
u(a,\phi)=f\left(\cos\left(\frac{\sqrt{6}}{4}\phi\right)a^{\frac{3}{2}}\right)\,,
\end{equation} 
\begin{equation}\label{new_coordinates_v}
v(a,\phi)=\frac{2}{3B}\sin\left(\frac{\sqrt{6}}{4}\phi\right)a^{\frac{3}{2}} + g\left(\cos\left(\frac{\sqrt{6}}{4}\phi\right)a^{\frac{3}{2}}\right)\,,
\end{equation} 
\end{subequations}
where $f$ and $g$ are arbitrary real functions of one variable. By setting $B=\frac{2}{3}$, $f(x)=x$ and $g(x)=0$, we have
\begin{subequations}\label{new_coordinates_final}
\begin{equation}\label{new_coordinates_final_u}
u(a,\phi)=\cos\left(\frac{\sqrt{6}}{4}\phi\right)a^{\frac{3}{2}}\,,
\end{equation} 
\begin{equation}\label{new_coordinates_final_v}
v(a,\phi)=\sin\left(\frac{\sqrt{6}}{4}\phi\right)a^{\frac{3}{2}} \,.
\end{equation} 
\end{subequations}
Inverting functions \eqref{new_coordinates_final}, we get the relations
\begin{subequations}\label{inverted_coordinates}
\begin{equation}\label{a}
a=\left(u^2+v^2\right)^{\frac{1}{3}}\,,
\end{equation}
\begin{equation}\label{phi}
\phi=\frac{4}{\sqrt{6}}\arctan\left(\frac{v}{u}\right)\,.
\end{equation}
\end{subequations}
Eqs. \eqref{inverted_coordinates} allow us to express the Lagrangian \eqref{new_Lagrangianbis} in the new coordinates as
\begin{equation}\label{new_Lagrangianbis_new_coordinates}
{\cal L}(u,v,\dot u,\dot v) = \frac{4}{3}\left({\dot u}^2 + {\dot v}^2\right) + V_0\/u^2 +M\/u\,.
\end{equation}
The Lagrange equations generated by the Lagrangian \eqref{new_Lagrangianbis_new_coordinates} are given by
\begin{subequations}\label{Lagrange_equations}
\begin{equation}\label{Lagrange_equation_u}
\frac{8}{3}\ddot u - 2V_0\/u -M=0\,,
\end{equation}
\begin{equation}\label{Lagrange_equation_v}
\frac{8}{3}\ddot v =0\,.
\end{equation}
\end{subequations}
Eqs. \eqref{Lagrange_equations} admit solutions of the form
\begin{subequations}\label{EL-solutions}
\begin{equation}\label{EL-solutions_u}
u(t)= C_1\/e^{\frac{\sqrt{3V_0}t}{2}} + C_2\/e^{-\frac{\sqrt{3V_0}t}{2}} - \frac{M}{2V_0}\,,
\end{equation}
\begin{equation}\label{EL-solutions_v}
v(t)=v_1\/t + v_0 \,.
\end{equation}
\end{subequations}
Similar solutions are obtained by choosing $G(x)=x^2$ in Eq. \eqref{solution_Fbis}. In this case, the point--like Lagrangian is expressed as 
\begin{equation}\label{new_Lagrangiantris}
{\cal L}= 3a{\dot a}^2 + a^3\left(\frac{{\dot\phi}^2}{2} + V_0\/\cos^2\left(\frac{\sqrt{6}}{4}\phi\right)\right) + M\cos^2\left(\frac{\sqrt{6}}{4}\phi\right)a^3\,.
\end{equation}
Performing the change of coordinates \eqref{inverted_coordinates},  Lagrangian \eqref{new_Lagrangiantris} assumes the form
\begin{equation}\label{new_Lagrangiantris_new_coordinates}
{\cal L}(u,v,\dot u,\dot v) = \frac{4}{3}\left({\dot u}^2 + {\dot v}^2\right) + \left(V_0+M\right)\/u^2\,.
\end{equation}
Lagrangian \eqref{new_Lagrangiantris_new_coordinates} yields the Euler-- Lagrange equations
\begin{subequations}\label{Lagrange_equations_tris}
\begin{equation}\label{Lagrange_equation_u_tris}
\frac{4}{3}\ddot u - \left(V_0+M\right)\/u =0\,,
\end{equation}
\begin{equation}\label{Lagrange_equation_v_tris}
\frac{8}{3}\ddot v =0\,.
\end{equation}
\end{subequations}
Eqs. \eqref{Lagrange_equations_tris} give solutions of the form
\begin{subequations}\label{EL-solutions_tris}
\begin{equation}
\label{EL-solutions_u_tris_a}
u(t)= C_1\/e^{\frac{\sqrt{3\left(V_0+M\right)}t}{2}} + C_2\/e^{-\frac{\sqrt{3\left(V_0+M\right)}t}{2}} \,,
\end{equation}
\begin{equation}
\label{EL-solutions_v_tris_b}
v(t)=v_1\/t + v_0 \,.
\end{equation}
\end{subequations}
In particular, for $V_0=-M$, we have
\begin{equation}\label{EL-solutions_u_tris_particular}
u(t)= u_1\/t + u_0\,.
\end{equation}
More in general, we can chose $G(x)=x^n$ in Eq. \eqref{solution_Fbis}. The corresponding point--like Lagrangian is given by
\begin{equation}\label{new_Lagrangian_general}
{\cal L}= 3a{\dot a}^2 + a^3\left(\frac{{\dot\phi}^2}{2} + V_0\/\cos^2\left(\frac{\sqrt{6}}{4}\phi\right)\right) + M\cos^n\left(\frac{\sqrt{6}}{4}\phi\right)a^{\frac{3n}{2}}\,.
\end{equation}
In the new coordinates \eqref{inverted_coordinates}, Lagrangian \eqref{new_Lagrangian_general} is expressed as
\begin{equation}\label{new_Lagrangian_general_new_coordinates}
{\cal L}(u,v,\dot u,\dot v) = \frac{4}{3}\left({\dot u}^2 + {\dot v}^2\right) + V_0\/u^2 + M\/u^n\,.
\end{equation}
The latter gives rise to Euler--Lagrange equations of the form
\begin{subequations}\label{Lagrange_equations_general}
\begin{equation}\label{Lagrange_equation_u_general}
\frac{8}{3}\ddot u - 2V_0\/u - nM\/u^{(n-1)}=0\,,
\end{equation}
\begin{equation}\label{Lagrange_equation_v_general}
\frac{8}{3}\ddot v =0\,.
\end{equation}
\end{subequations}
Eqs. \eqref{Lagrange_equations_general} admit the following solutions
\begin{subequations}\label{EL-solutions_general}
\begin{equation}\label{EL-solutions_u_general}
\pm\int\frac{2du}{\sqrt{3V_0\/u^2 + 3nM\/u^n + C_1}}= t+C_2\,,
\end{equation}
\begin{equation}\label{EL-solutions_v_general}
v(t)=v_1\/t + v_0 \,.
\end{equation}
\end{subequations}
Similar solutions can be obtained by setting $G(x)=e^x$, $G(x)=ln(x)$ or $G(x)=\sqrt{x}$ in Eq. \eqref{solution_Fbis}. For $G(x)=e^x$, the corresponding point--like Lagrangian is given by
\begin{equation}\label{new_Lagrangian_general2}
{\cal L}= 3a{\dot a}^2 + a^3\left(\frac{{\dot\phi}^2}{2} + V_0\/\cos^2\left(\frac{\sqrt{6}}{4}\phi\right)\right) + M\/e^{cos\left(\frac{\sqrt{6}}{4}\phi\right)a^{\frac{3}{2}}}\,.
\end{equation}
In the new coordinates \eqref{inverted_coordinates},  Lagrangian \eqref{new_Lagrangian_general2} is expressed as
\begin{equation}\label{new_Lagrangian_general_new_coordinates2}
{\cal L}(u,v,\dot u,\dot v) = \frac{4}{3}\left({\dot u}^2 + {\dot v}^2\right) + V_0\/u^2 + M\/e^u\,.
\end{equation}
From Eqs. \eqref{new_Lagrangian_general_new_coordinates2}, we derive Lagrange equations of the form
\begin{subequations}\label{Lagrange_equations_general2}
\begin{equation}\label{Lagrange_equation_u_general2}
\frac{8}{3}\ddot u - 2V_0\/u - M\/e^u=0\,,
\end{equation}
\begin{equation}\label{Lagrange_equation_v_general2}
\frac{8}{3}\ddot v =0\,.
\end{equation}
\end{subequations}
Eqs. \eqref{Lagrange_equations_general2} admit the following solutions
\begin{subequations}\label{EL-solutions_general2}
\begin{equation}\label{EL-solutions_u_general2}
\pm\int\frac{2du}{\sqrt{3V_0\/u^2 + 3M\/e^u + C_1}}= t+C_2\,,
\end{equation}
\begin{equation}\label{EL-solutions_v_general2}
v(t)=v_1\/t + v_0 \,.
\end{equation}
\end{subequations}
For $G(x)=ln(x)$, the point--like Lagrangian is
\begin{equation}\label{new_Lagrangian_general3}
{\cal L}= 3a{\dot a}^2 + a^3\left(\frac{{\dot\phi}^2}{2} + V_0\/\cos^2\left(\frac{\sqrt{6}}{4}\phi\right)\right) + M\/ln\left(cos\left(\frac{\sqrt{6}}{4}\phi\right)a^{\frac{3}{2}}\right)\,.
\end{equation}
In the coordinates \eqref{inverted_coordinates}, Lagrangian \eqref{new_Lagrangian_general3} is expressed as
\begin{equation}\label{new_Lagrangian_general_new_coordinates3}
{\cal L}(u,v,\dot u,\dot v) = \frac{4}{3}\left({\dot u}^2 + {\dot v}^2\right) + V_0\/u^2 + M\/ln(u)\,.
\end{equation}
The induced Lagrange equations are of the form
\begin{subequations}\label{Lagrange_equations_general3}
\begin{equation}\label{Lagrange_equation_u_general3}
\frac{8}{3}\ddot u - 2V_0\/u - \frac{M}{u}=0\,,
\end{equation}
\begin{equation}\label{Lagrange_equation_v_general3}
\frac{8}{3}\ddot v =0\,.
\end{equation}
\end{subequations}
Solutions of \eqref{Lagrange_equations_general3} are
\begin{subequations}\label{EL-solutions_general3}
\begin{equation}\label{EL-solutions_u_general3}
\pm\int\frac{2du}{\sqrt{3V_0\/u^2 + 3M\/ln(u) + C_1}}= t+C_2\,,
\end{equation}
\begin{equation}\label{EL-solutions_v_general3}
v(t)=v_1\/t + v_0\,. 
\end{equation}
\end{subequations}
Finally, for $G(x)=\sqrt{x}$, the point--like Lagrangian is given by
\begin{equation}\label{new_Lagrangian_general4}
{\cal L}= 3a{\dot a}^2 + a^3\left(\frac{{\dot\phi}^2}{2} + V_0\/\cos^2\left(\frac{\sqrt{6}}{4}\phi\right)\right) + M\/\sqrt{\left(cos\left(\frac{\sqrt{6}}{4}\phi\right)a^{\frac{3}{2}}\right)}\,.
\end{equation}
In the coordinates \eqref{inverted_coordinates}, Lagrangian \eqref{new_Lagrangian_general4} is expressed as
\begin{equation}\label{new_Lagrangian_general_new_coordinates4}
{\cal L}(u,v,\dot u,\dot v) = \frac{4}{3}\left({\dot u}^2 + {\dot v}^2\right) + V_0\/u^2 + M\/\sqrt{u}
\end{equation}
and the corresponding Euler--Lagrange equations are
\begin{subequations}\label{Lagrange_equations_general4}
\begin{equation}\label{Lagrange_equation_u_general4}
\frac{8}{3}\ddot u - 2V_0\/u - \frac{M}{2\sqrt{u}}=0\,,
\end{equation}
\begin{equation}\label{Lagrange_equation_v_general4}
\frac{8}{3}\ddot v =0\,.
\end{equation}
\end{subequations}
Eqs. \eqref{Lagrange_equations_general4} admit the following solutions
\begin{subequations}\label{EL-solutions_general4}
\begin{equation}\label{EL-solutions_u_general4}
\pm\int\frac{2du}{\sqrt{3V_0\/u^2 + 3M\/\sqrt{u} + C_1}}= t+C_2\,,
\end{equation}
\begin{equation}\label{EL-solutions_v_general4}
v(t)=v_1\/t + v_0\,. 
\end{equation}
\end{subequations}
 As final remark, we observe that, in order to solve  Eq. (\ref{Noether_equations_4bis})\,- or, analogously,   Eq. \eqref{Noether_equations_4}\,-\,, it is possible to redefine $F(a,\phi)$  to absorb all terms, including the potential $V(\phi)$, into a single function:  
 \begin{equation}
  F(a,\phi) \rightarrow \tilde{F}(a,\phi) = a^3 V(\phi) + M a^{-3 (\gamma-1)}(1+F(a,\phi) )\,.
 \label{redefinition-REf}
 \end{equation}
  It turns out that,  using Eqs. \eqref{solutions_alpha_betabis} with $A=0$,
  \begin{eqnarray}
&& F(a,\phi ) = \frac{a^{3 \gamma -3} \left(M c_1\left[\sqrt{\frac{2}{3}}
   \left(2 \log \left(\sec \left(\frac{1}{2} \sqrt{\frac{3}{2}} \phi \right)\right)-3
   \log (a)\right)\right]\right)}{M} \nonumber 
\\ && -\frac{a^{3 \gamma} V(\phi )-3 a^{3 \gamma -2} k}{M}-1\,,
   \end{eqnarray}
where $c_1[x]$ is an arbitrary function of one real variable.  For instance, by choosing
\begin{eqnarray}
&&c_1\left[\sqrt{\frac{2}{3}} \left(2 \log \left(\sec \left(\frac{1}{2} \sqrt{\frac{3}{2}}
   \phi \right)\right)-3 \log (a)\right)\right] =\\ \nonumber
 &&  \exp \left(-\frac{1}{3} m \left(2
   \log \left(\sec \left(\frac{1}{2} \sqrt{\frac{3}{2}} \phi \right)\right)-3 \log
   (a)\right)\right)\,,
\end{eqnarray}
we have the function $F(a,\phi)$ in the form: 
\begin{equation}
F(a,\phi)=\frac{a^{3 \gamma -3} \left(M a^m \sec ^{-\frac{2 m}{3}}\left(\frac{1}{2}
   \sqrt{\frac{3}{2}} \phi \right)-a^3 V(\phi )-3 a k\right)}{M}-1\,.
   \label{solutionFgen}
\end{equation}
Inserting Eq. \eqref{solutionFgen} into Eq. \eqref{Lagrangianbis}, Lagrangian \eqref{Lagrangianbis} assumes the expression
\begin{equation}\label{new_LagrangianFgen}
{\cal L}=a^m \sec ^{-\frac{2 m}{3}}\left(\frac{1}{2} \sqrt{\frac{3}{2}} \phi \right)+\frac{a^3
  {\dot \phi}^2}{2}+3 a  {\dot a}^2.
  \end{equation}
It turns out that this choice in Eq. (\ref{redefinition-REf}) has no effect in the search for Noether symmetries and the results are the same with only a single function being involved. However, the form used in the action (\ref{Lagrangianbis}) offers the advantage to highlight  contributions from different terms (scalar field potential, matter etc.). 

\section{Exact solutions as dark energy: comparison between theoretical prediction and observations}
\label{Sec4}
 Let us  show now that some of the above solutions  can reproduce, both for the standard ($\epsilon=1$) and  the phantom scalar field ($\epsilon= -1$),  the accelerated expansion of the Universe. In particular,  they are compatible with different observational datasets, as the SNeIa Pantheon, the gamma ray bursts Hubble diagram, and  measurements of the Hubble parameter. 
\subsection{The case of the standard scalar field }
For our purposes,  we consider the case described by the Eqs. \eqref{solution_u} and \eqref{solution_v} with $h=\displaystyle\frac{3}{2}$. It turns out that 
\begin{eqnarray}
&& u(t)=\frac{3 \Sigma  t}{4}+u_0 \\ && v(t)= \frac{3}{8} t^2 \left(M Q+2 u_0 V_0\right)+\frac{t u_0 \left(M Q+u_0
   V_0\right)}{\Sigma }+\frac{3}{16} \Sigma  t^3 V_0\,.\nonumber
\end{eqnarray}
The scale factor and the scalar field can be expressed as functions of  $u(t)$ and $v(t)$. Actually $a(t)=\left(uv\right)^{1/3}$, and $\phi(t)= -\sqrt{2/3}\ln{\frac{u}{v}}$.
Moreover we impose the condition $a(0)=0$, and we set  
the  age of the Universe, $t_0$, as time-scale ($t_0 = 1$). Therefore  the 
expansion rate $H(t)$ is dimensionless, and its actual value ${H}_0=H(t_0)$ 
is clearly of order $1$. This means that it is numerically different from the Hubble constant  usually measured 
in ${\rm km s^{-1} Mpc^{-1}}$. Actually, $H_0$ 
depends on the integration constants. We then set $a_0 = a(1) = 1$, and $H_0=H(1)$.  
These conditions induce some constraints among the integration constants:
for instance from the condition
$a(t_0=1)=1$ we obtain
\begin{equation}\label{V0formula}
V_0= \frac{64 \Sigma -2 M Q \left(3 \Sigma +4 u_0\right) \left(3 \Sigma +8 u_0\right)}{\left(3 \Sigma +4
   u_0\right) \left(3 \Sigma ^2+12 \Sigma  u_0+16 u_0^2\right)}\,,
\end{equation}
and from $H(t_0=1)=H_0$ we obtain-once that we substitute $V_0$ with the relation in Eq. \eqref{V0formula}:
\begin{eqnarray}\label{Qformula}
&&Q_0=-\left\{32 \left(9 \left(3 H_0-4\right) \Sigma ^3+144 \left(H_0-1\right) \Sigma ^2 u_0+96 \left(3 H_0-2\right)
   \Sigma  u_0^2+64 \left(3 H_0-1\right) u_0^3\right)\right\}\times\nonumber\\
&&   \left\{3 M \left(\Sigma +4 u_0\right) \left(3 \Sigma +4
   u_0\right){}^3\right\}^{-1}\,.
\end{eqnarray}

\begin{figure}[!ht]
\includegraphics[width=\linewidth,clip]{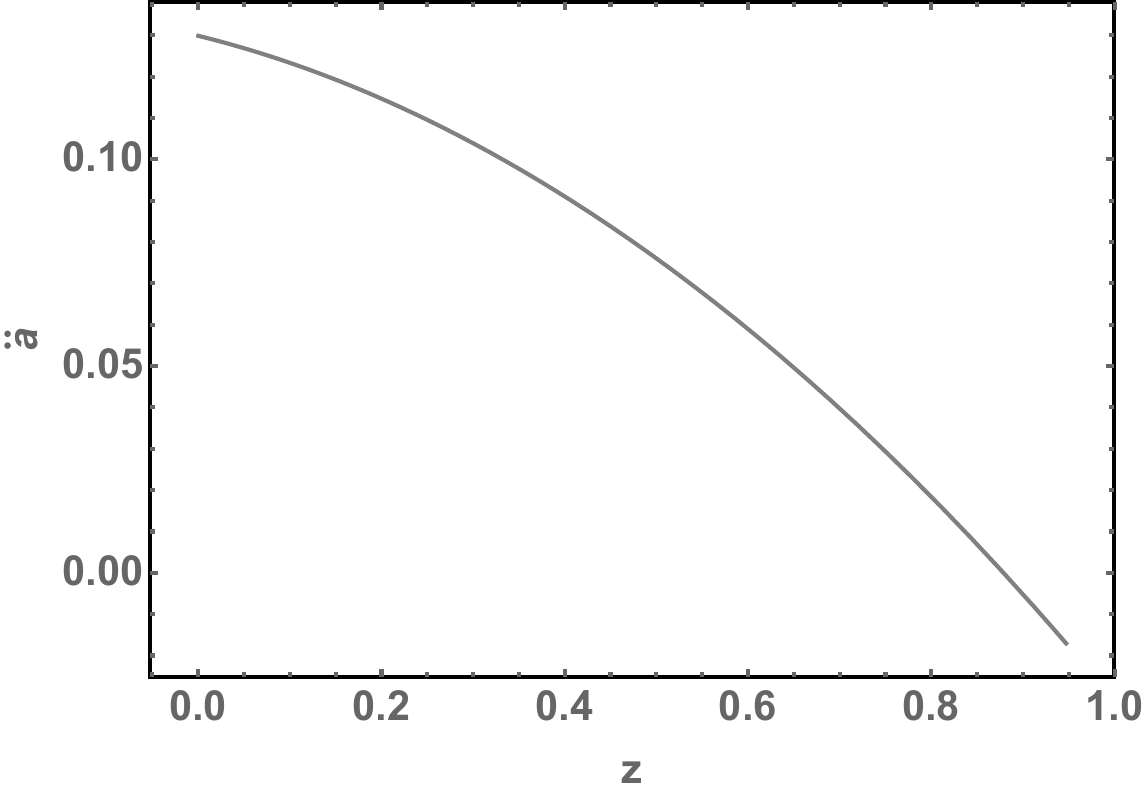}
\caption{The redshift dependence of the acceleration for the standard scalar field model, corresponding to fixed values  of $H_0=1$, $u_0=5.12$ and $\Sigma=1.2$. The model provides an accelerated expansion and  the transition to the decelerated regime occurs at a  redshift compatible with  observations.}
\label{acc-stfield}
\end{figure}

By means of these choices the scale factor and the scalar field are parametrized by $H_0$, $u_0$ and $\Sigma$:
\begin{eqnarray}
\label{acubestandard}
a^3(t)&=& \dfrac{3 \Sigma ^2 t^4 \left(27 \left(H_0-1\right) \Sigma ^2+36 \left(3 H_0-2\right) \Sigma 
   u_0+32 \left(3 H_0-1\right) u_0^2\right)}{\left(\Sigma +4 u_0\right) \left(3 \Sigma +4 u_0\right){}^3}+\nonumber \\
&& \dfrac{4 t^3 \left(\frac{27}{4} \left(4-3 H_0\right) \Sigma ^4+72 \left(3 H_0-2\right) \Sigma
   ^2 u_0^2+80 \left(3 H_0-1\right) \Sigma  u_0^3\right)}{\left(\Sigma +4 u_0\right) \left(3 \Sigma +4 u_0\right){}^3}+\nonumber\\
   && \dfrac{4 t^2 u_0 \left(27 \left(4-3 H_0\right) \Sigma ^3-216 \left(H_0-1\right) \Sigma ^2
   u_0+64 \left(3 H_0-1\right) u_0^3\right)}{\left(\Sigma +4 u_0\right) \left(3 \Sigma +4 u_0\right){}^3}+\nonumber\\&&
   \dfrac{32 t u_0^2 \left(3 \left(4-3 H_0\right) \Sigma ^2-30 \left(H_0-1\right) \Sigma  u_0+8
   \left(2-3 H_0\right) u_0^2\right)}{\left(\Sigma +4 u_0\right) \left(3 \Sigma +4 u_0\right){}^3}\,,
    \end{eqnarray}
    \begin{eqnarray}
\label{exphistandard}
\displaystyle e^{\sqrt{\frac{3}{2}} \phi}&=&\dfrac{v(t)}{u(t)}=\left(\frac{3 \Sigma t}{4} + u_0\right)^{-1}\times \\ && \left\{\dfrac{ 4 \Sigma  t^3 \left(27 \left(H_0-1\right) \Sigma ^2+36 \left(3 H_0-2\right) \Sigma 
u_0+32 \left(3 H_0-1\right) u_0^2\right)}{\left(\Sigma +4 u_0\right) \left(3 \Sigma +4 u_0\right)^3}+ \right.\nonumber \\ 
&& \left. \dfrac{4 t^2 \left(9 \left(4-3 H_0\right) \Sigma^3-36 \left(H_0-1\right) \Sigma ^2 u_0+48 \left(3 H_0-2\right) \Sigma  u_0^2+64 \left(3 H_0-1\right) u_0^3\right)}{\left(\Sigma +4 u_0\right) \left(3 \Sigma +4 u_0\right)^3}\right.\nonumber \\ 
&& \left. \dfrac{32 t u_0 \left(3 \left(4-3 H_0\right) \Sigma ^2-30 \left(H_0-1\right) \Sigma  u_0+8  \left(2-3 H_0\right) u_0^2\right)}{\left(\Sigma +4 u_0\right) \left(3 \Sigma +4 u_0\right)^3}\right\}\,.
\end{eqnarray}
This exact solution provides an accelerated expansion  as shown in Fig. \ref{acc-stfield}.

\begin{figure}[!ht]
\includegraphics[width=\linewidth,clip]{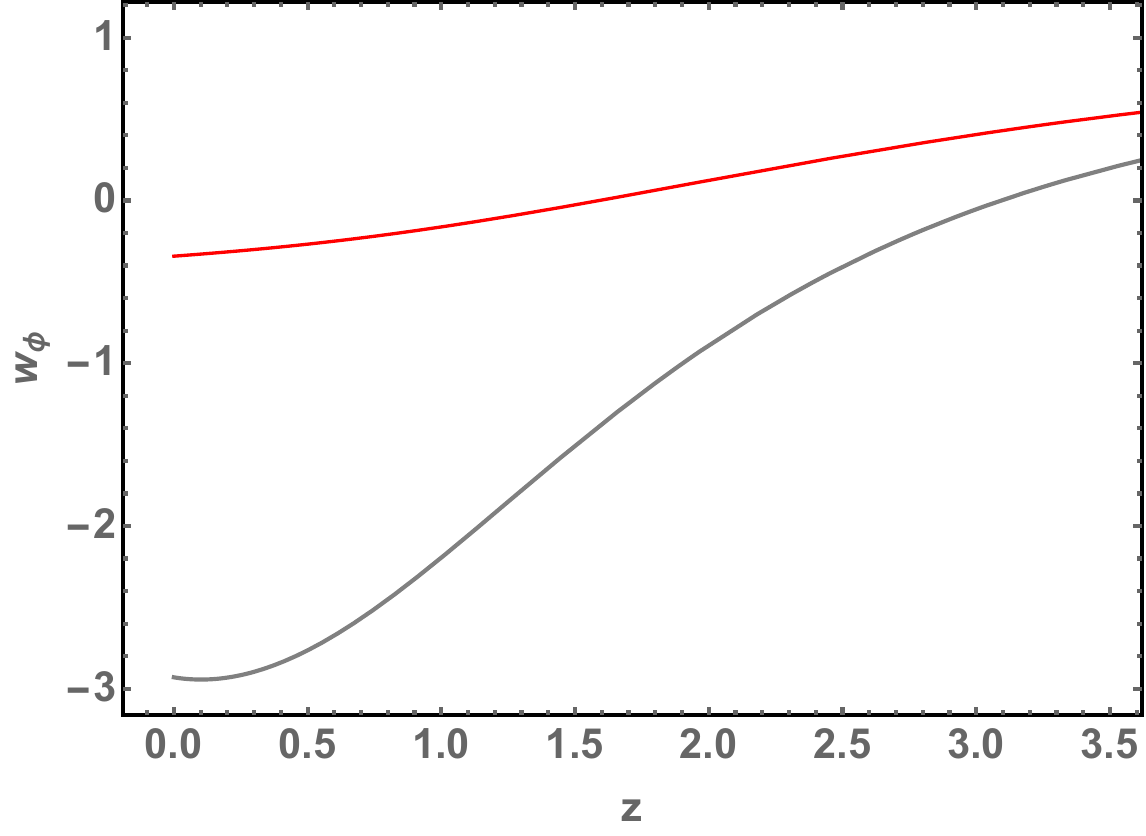}
\caption{The redshift dependence of the equation of state parameter $w^{eff}_{\phi}$ (grey line) and $w_{\phi}$ (red line)  for the standard scalar field model, corresponding to fixed value of $H_0=1$, $u_s0=5.12$ and $\Sigma=1.2$. The values of the parameters are chosen to highlight the different behavior between the two function: it is  evident the  {\it super-quintessential} nature of the equation of state ($w^{eff}_{\phi}  <1$) due to the interaction term.}
\label{eos-stfield}
\end{figure}

Moreover, using the analytical expressions for $a(t)$ and $\phi(t)$,  we can construct the standard quantities $\rho_{\phi}$,$p_{\phi}$,$V_{\phi}$, $w_{\phi}$,  and the effective quantities $\rho^{eff}_{\phi}$, $p^{eff}_{\phi}$, and $w^{eff}_{\phi}$, defined by  Eqs. \eqref{fi-stdensity},\eqref{fi-effdensity},\eqref{fi-pressure},\eqref{eos-eff} with $\epsilon=1$. In Fig. \ref{eos-stfield}, we compare the redshift behaviour of $w_{\phi}$ and $w^{eff}_{\phi}$ for some values of the parameters. 
\subsection{Supernovae and GRB Hubble diagram}
\label{Sec4b}
The SNIa Hubble diagram provided the first strong evidence of the present accelerating expansion of the
Universe. Here we consider the  Pantheon compilation, consisting of $1048$ SNIa in the range $0.01 < z < 2.26$. This sample combines  $365$ spectroscopically confirmed SNIa, discovered by the Pan-STARRS1 PS1 Medium Deep Survey, the subset of $279$ PS$1$ SNIa in the range ($0.03 < z < 0.68$), distance estimates from SDSS, SNLS, and various low redshift and HST samples \cite{scolnic2018}.
The SNIa observations provide the apparent magnitude $m(z)$, related to the Hubble free luminosity distance through the relation:
\begin{equation}
m_{th}(z)={\bar M}+ 5 \log_{10} (D_L (z))\,. \label{mdl}
\end{equation}
Here ${\bar M}$ is the zero point offset and depends on the absolute magnitude
$M$ and on the Hubble parameter. The theoretical distance modulus is therefore defined as
\begin{equation}
\label{lumdist}
\mu_{th}(z_i,\{\theta_{p}\}) =5 \log_{10} (D_L(z_i,\{\theta_{p}\})) +\nu_0\,,
\end{equation}
where $D_L$ is the luminosity distance:
\begin{equation}
D_L =\frac{c}{100 h} (1+z)\int^{z}_{0}\frac{1}{H(\zeta, \theta)}d\zeta\,. 
\end{equation}
The parameter $\nu_0$ in Eq. \eqref{lumdist} encodes the Hubble constant. The absolute magnitude $M$ and$\{\theta_{p}\}$ are the parameters of the model. Actually, it is well known that, using only SNeIa, one cannot constrain the Hubble constant, without including measurements of its local value, since this is degenerate with $M$. 

\begin{figure}[!ht]
\includegraphics[width=\linewidth,clip]{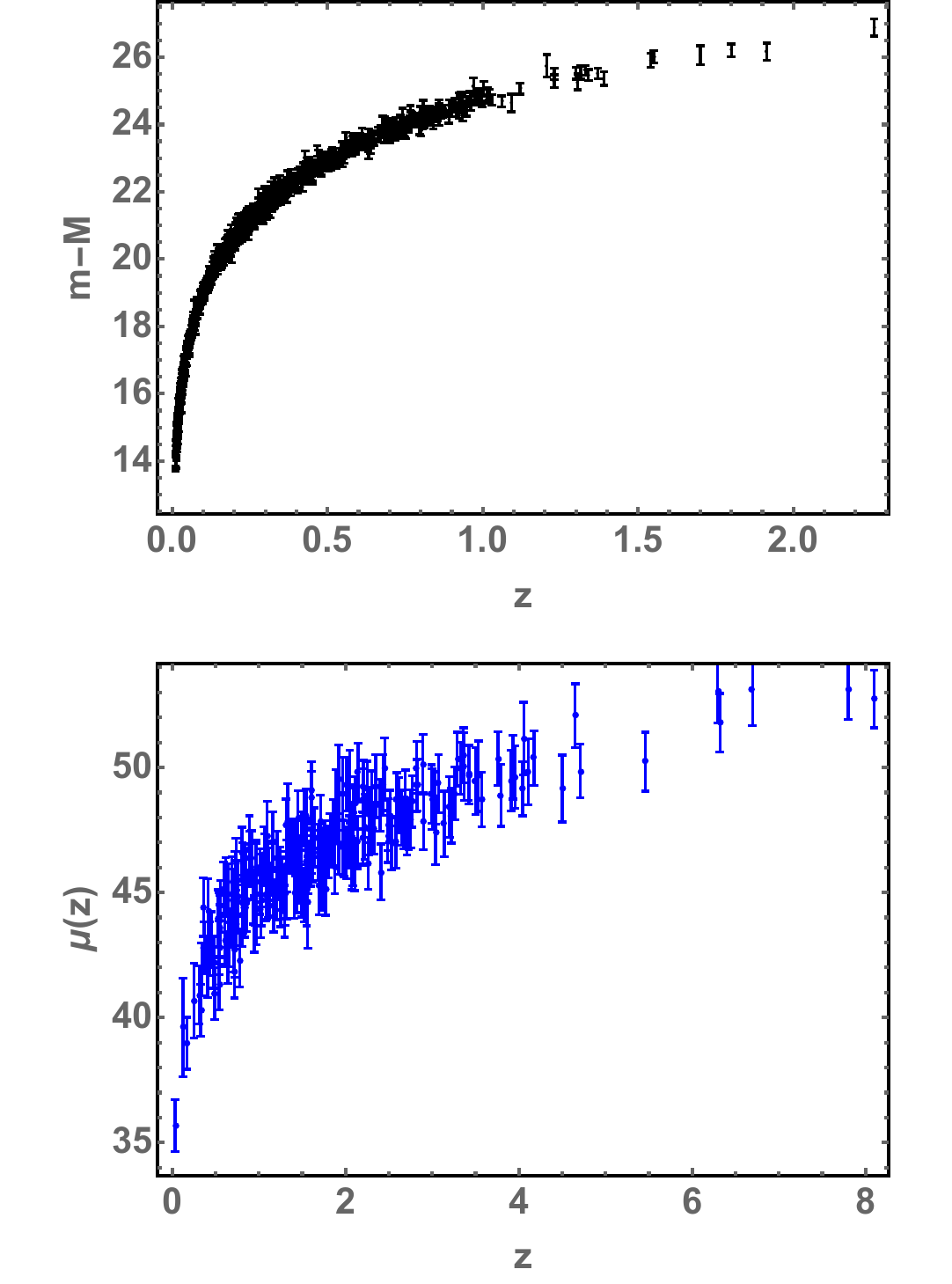}
\caption{The Hubble diagram of SNIa (upper panel) and GRBs (lower panel), with their respective $1 \sigma$ uncertainties.}
\label{HDplot}
\end{figure}

Gamma-ray bursts (GRBs) are  the brightest cosmological sources in the Universe, thanks to the enormous amount of energy released  in  tens or hundreds of
seconds: actually the isotropic radiated energy, $E_{iso}$, can reach  $10^{54}$ erg.  Moreover their redshift distribution extends up to  $ z\sim  9.4$: therefore they are good candidates for  cosmological investigation. Unfortunately GRBs  are not standard candles, since their peak luminosity spans a wide range. However it is possible to consider them as distance indicators calibrating some empirical correlations of distance-dependent quantities and rest-frame observables\citep{Amati02,Tanvir,LR22,grb-quasar}. Here we consider the GRB Hubble diagram built up from the the $E_{\rm p,i}$-$ E_{\rm iso }$  correlation. Actually, it is well known that GRBs have non-thermal spectra modeled by a smoothly broken power law with  two indices (a low index $\alpha$, and a high index $\beta$ ), named the band function,$N(E)$. Their spectra show a peak corresponding the photon energy $E_{\rm p} = E_0 (2 + \alpha)$. Moreover, for GRBs with measured spectrum and redshift, it is possible to evaluate the intrinsic peak energy, $E_{\rm p,i} = E_{\rm p} (1 + z)$ and the isotropic equivalent radiated energy, defined as: 
\begin{equation}
E_{\rm iso}= 4 \pi D_L(z,{\rm \theta}) \left(1+z\right)^{-1}\int^{10^4/(1+z)}_{1/(1+z)} E N(E)\,,
dE\,,
\label{eqEiso}
\end{equation}
where 
\[N(E)=\left\{
\begin{array}{ll}
 A \left(\frac{E}{100keV}\right)^{\alpha} \exp{\left(-{\frac{E}{E_0}}\right)} & \left(\alpha-\beta\right)E_0\geq E \;\; ,\\
 A \left(\frac{\left(\alpha-\beta\right)E}{100keV}\right)^{\alpha-\beta} \exp{\left(\alpha-\beta\right)\left(\frac{E}{100keV}\right)^{\beta}} & \left(\alpha-\beta\right)E_0\leq E \;\; .\\
\end{array}
\right. \]
The existence of a  correlation between \epi and \eiso for long GRBs was discovered in 2002 \citep{Amati02}, and was confirmed by later measurements by several different GRB detectors. It can be modeled as a linear relation between the logarithms of the two quantities:

\begin{equation}
\log \left[\frac{E_{\rm{p, i}}}{\rm{keV}}\right]=b+a \log \left[\frac{E_{\rm iso}}{10^{52}\;\rm{erg}}\right] \;\; ,
 \label{correlation_amati}
\end{equation}
The \epeiso~correlation  is characterized by an intrinsic additional extra-Poissonian scatter, $\sigma_{int}$, around the best-fit line that has to be taken into account and determined together with $(a, b)$ by the fitting procedure. After that  values $(a, b)$ are estimated, it is possible to obtain the energy $E_{\rm iso}$ of each burst at high redshift  through Eq.~\ref{correlation_amati}, and the luminosity distance, $D_{\rm L}(z)$ from Eq. \eqref{eqEiso}, building up the GRB Hubble diagram. Here we use the GRB Hubble diagram presented in \citep{LR22},  \citep{2021MNRASD}, and in  \citep{cosmo_ester}. 
In Fig. \ref{HDplot},  we show the SNIa and GRB Hubble diagram.
\subsection{Direct H(z) measurements}
\begin{figure}[!ht]
\includegraphics[width=\linewidth,clip]{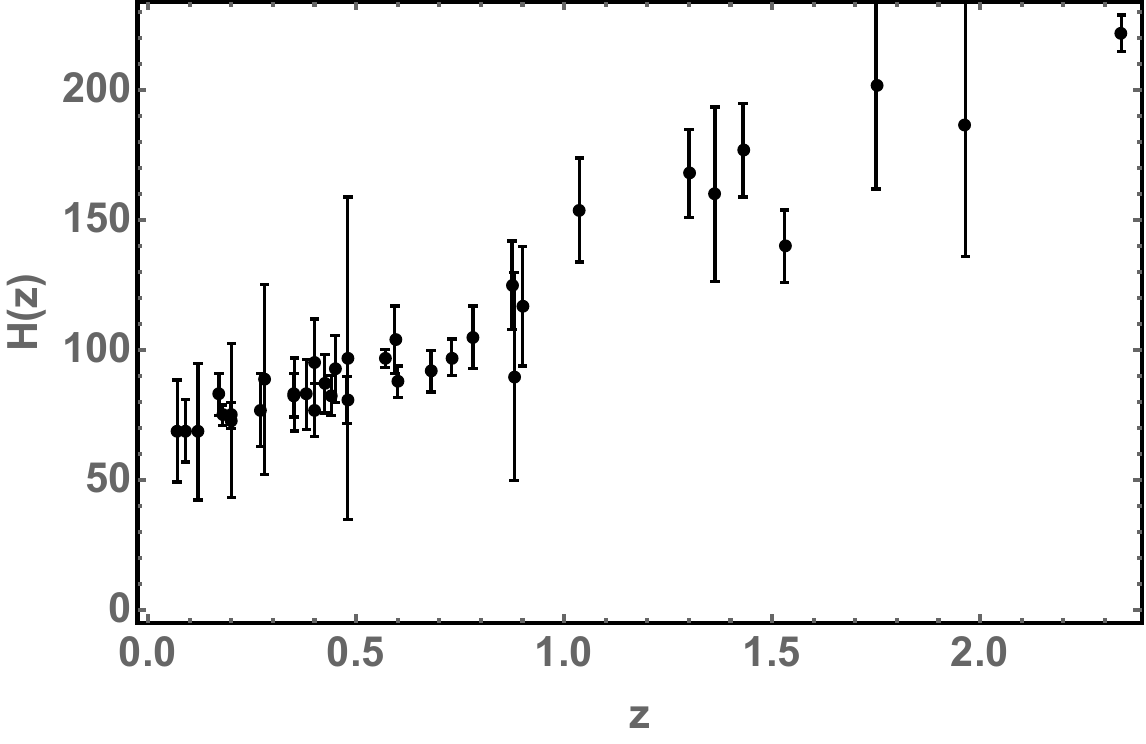}
\caption{The direct $H(z)$ measurements used in our analysis. }
\label{Hzplot}
\end{figure}

The accurate and direct determination of the expansion rate of the Universe,  $H(z)$, has become one of the main drivers in precision cosmology, since it can provide fundamental information about  the possible physical mechanisms underlying the late time acceleration. The Hubble parameter, defined as $H(z) =
\displaystyle \frac{\dot a}{a}$,  depends on the differential age of the Universe as a
function of redshift and can be measured  using the cosmic chronometers. The quantity $dz$ is obtained from spectroscopic
surveys with high resolution, and  the differential evolution of the age of the Universe $dt$ in the  redshift interval
$dz$ can be measured provided that  appropriate probes of the aging of the Universe, that is, just the cosmic chronometers, are
identified. The most reliable cosmic chronometers, observable at high redshift,  are old early-type galaxies  evolving passively on a
timescale much longer than their age difference.  These galaxies formed the majority of their stars rapidly and  early and they have not experienced  subsequent major star formation or merging  episodes. Moreover, the Hubble parameter can also be obtained from  BAO measurements, observing the typical acoustic scale in the light-of-sight direction. Here we used a list of direct $H(z)$
measurements in the redshift range $z\sim 0.07-2.3$, compiled in \citep{2016JCAPM}, and \citep{2016EPJCG}, as shown in Fig.\ref{Hzplot}
\subsection{Statistical analysis}
\begin{figure}[!ht]
\includegraphics[width=\linewidth,clip]{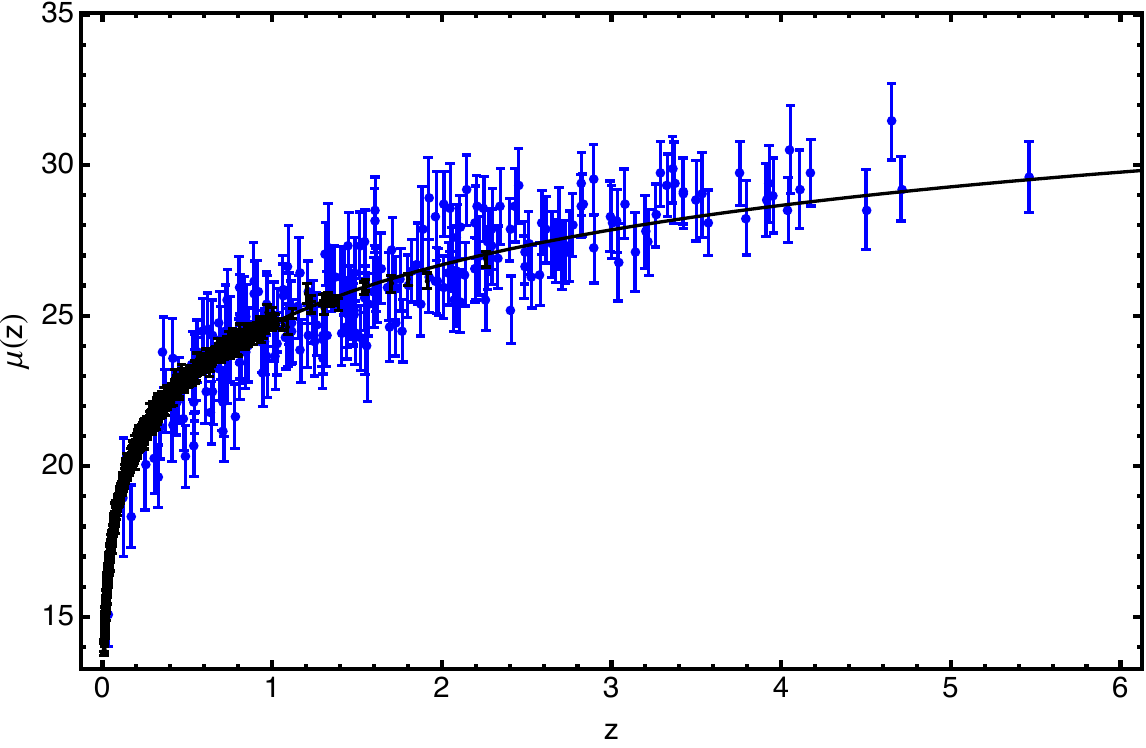}
\caption{Comparison between  GRB and Pantheon data vs  theoretical distance modulus,
corresponding to the best fit values of the parameters for the standard scalar field model. }
\label{hubbledatabf}
\end{figure}

\begin{figure}[!ht]
\includegraphics[width=\linewidth,clip]{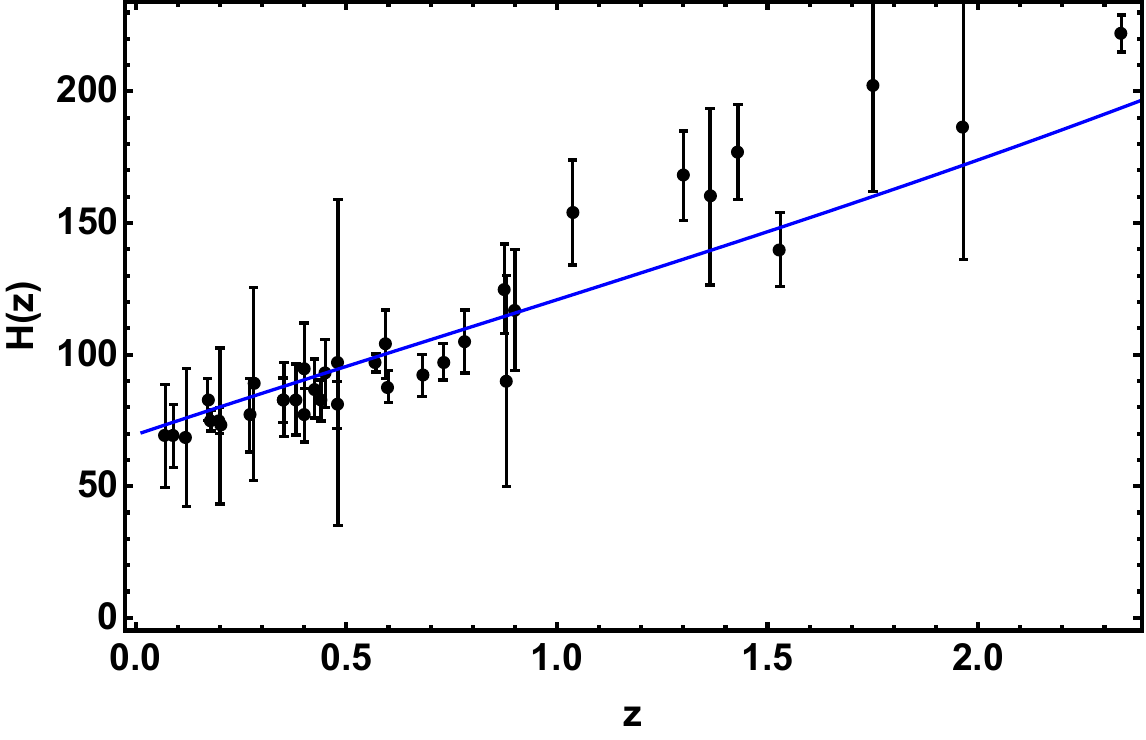}
\caption{Comparison between the $H(z)$ data vs the theoretical predictions,
corresponding to the best fit values of the parameters  for the standard scalar field model. }
\label{Hzdatabf}
\end{figure}

To test the cosmological model described above, we use a Bayesian approach based on the Markov Chain Monte Carlo (MCMC) method \citep{MCMC}. We set
the starting points for our chains performing a preliminary fit  to maximize the
likelihood function ${\cal{L}}({\bf p})$:
\begin{eqnarray}
\footnotesize
{\mathcal{L}}({\bf p}) & \propto & \frac {\exp{(-\chi^2_{SNIa/GRB}/2)}}{(2 \pi)^{\frac{{\cal{N}}_{SNIa/GRB}}{2}} |{\bf C}_{SNIa/GRB}|^{1/2}}  \nonumber\\ ~ & \times & \frac{\exp{(-\chi^2_{H}/2})}{(2 \pi)^{{\cal{N}}_{H}/2} |{\bf C}_{H}|^{1/2}} \,,
\label{defchiall}
\end{eqnarray}
where
\begin{equation}
\chi^2(\mathrm {\bf p}) = \sum_{i,j=1}^{N} \left( x_i -
x^{th}_i(\bf p)\right)C^{-1}_{ij}  \left( x_j - x^{th}_j(\bf
p)\right) \,. \label{eq:chisq}
\end{equation}
In Eq.\eqref{eq:chisq}, $\bf p$ indicates the parameters of the cosmological model, $N$ is the number of data points,  $\mathrm x_i$ is the $i-th$ measurement, and
$ x^{th}_i(\bf p)$ indicate the theoretical predictions.
$C_{ij}$ is the covariance matrix for the SNIa/GRB/H data. Moreover we used flat priors on the  parameters\,, and we apply the Gelman--Rubin test for the convergence of the five running chains.  We make thin the chains discarding the first $30\%$  of
 iterations at the beginning of any run, and we finally extract the best fit values and the regions of confidence on the parameters by co-adding
the thinned chains. In Table
\ref{tab1},  we present
the results of our analysis. In Figs. \eqref{hubbledatabf} and \eqref{Hzdatabf}, we plot  data vs  theoretical predictions.
\begin{table*}
\begin{center}
\resizebox{8cm}{!}{
\begin{tabular}{cccccc}
\, & \multicolumn{4}{c}{\bf Standard scalar field}   \\
\, & \, & \, & \, & \,   \\
\hline
\, & \, & \, & \, & \,    \\
$Id$ & $\langle x \rangle$ & $\tilde{x}$ & $68\% \ {\rm CL}$  & $95\% \ {\rm CL}$  \\
\hline \hline
\, & \, & \, & \, & \,  \\
\hline \, & \multicolumn{4}{c}{SNIa /GRBs/H(z)}  
 \\
\hline
\, & \, & \, & \, & \,  \\
$H_0$ &1.01 &1.02& (0.93, 1.08) & (0.90, 1.1)  \\
\, & \, & \, & \, & \, \\
$u_0$ &5.17& 4.2& (1.78,  8.3) & (1.07,  12.3) \\
\, & \, & \, & \, & \,   \\
$\Sigma$ &1.5&1.1 & (0.39\,,2.9) & (0.14\,, 4.6)  \\
\, & \, & \, & \, & \,  \\
$h$ &0.7& 0.69 & (0.63, 0.77) & (0.61, 0.79)   \\
\, & \, & \, & \, & \,  \\
\hline
\end{tabular}}
\end{center}
\caption{Constraints on the standard scalar field parameters from different data: combined SNIa  and  GRB Hubble diagrams,  and
$H(z)$ data sets. Columns show the mean $\langle x \rangle$ and median $\tilde{x}$ values  and the $68\%$ and $95\%$
confidence limits.}  \label{tab1}
\end{table*}

\subsection{The case of the phantom scalar field}
Also the phantom scalar field, selected by the  Noether symmetry, can provide a late accelerated expansion. Let us consider 
 the case described by the Eqs. \eqref{EL-solutions_u_tris_a} and \eqref{EL-solutions_v_tris_b}. It is  $a(t)=\left(u(t)^2+ v(t)^2\right)^{1/3}$, and $\phi(t)= \dfrac{4}{\sqrt 6} \arctan \left(\dfrac{v(t)}{u(t)}\right)$.
 
Also in this case we impose the condition $a(0)=0$, and we set the 
the  age of the Universe, $t_0$, as a time-scale ($t_0 = 1$). Indeed, we set $a_0 = a(1) = 1$, and $H_0=H(1)$.  
These conditions induce constraints among the integration constants: The scale factor and the scalar field are therefore parametrized by $H_0$,  and $\alpha $\,(where $\alpha=\sqrt{3\left(V_0+M\right)}$).

\begin{figure}[!ht]
\includegraphics[width=\linewidth,clip]{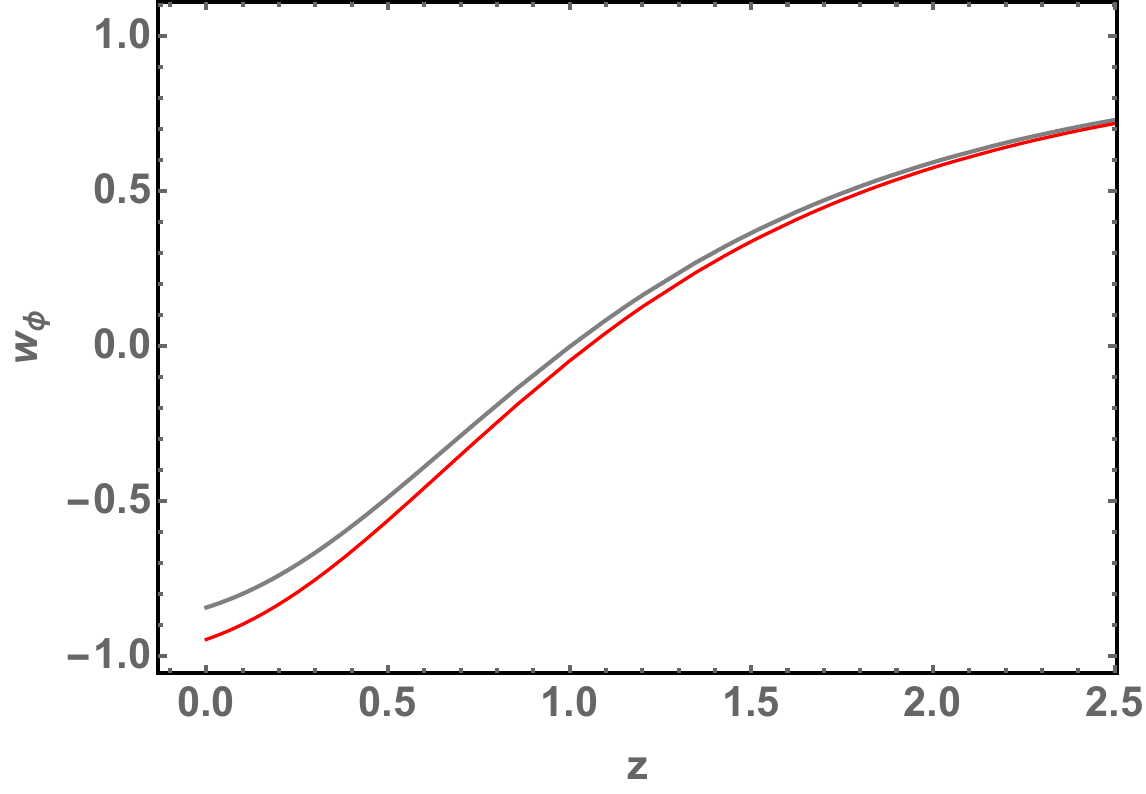}
\caption{The redshift dependence of the equation of state parameter $w^{eff}_{\phi}$ (red line) and $w_{\phi}$ (gray line) for the phantom scalar field model, corresponding to fixed values of $H_0=0.98$, $\alpha=3.2$ and $\Omega_m=0.27$.}
\label{eos-phinteractingfield}
\end{figure}

By the analytical expressions for $a(t)$ and $\phi(t)$,  we can construct the standard quantities $\rho_{\phi}$,$p_{\phi}$,$V_{\phi}$, $w_{\phi}$,  and the effective quantities $\rho^{eff}_{\phi}$, $p^{eff}_{\phi}$, and $w^{eff}_{\phi}$, defined by  Eqs. \eqref{fi-stdensity},\eqref{fi-effdensity},\eqref{fi-pressure},\eqref{eos-eff} with $\epsilon=-1$. It is worth noticing that in order to evaluate the interaction contribution to the effective quantities, we parametrize the present matter density, $M$, in terms of the associated $\Omega_M =\dfrac{M}{3 H_0^2}$.  In Fig. \ref{eos-phinteractingfield}, we compare the redshift behaviour of $w_{\phi}$ and $w^{eff}_{\phi}$ for some values of the parameters. In order to test this model, we use the same data-samples, and we apply the same statistical analysis described above. In  Table \ref{tab2},  we present the results of our analysis.
\begin{table*}
\begin{center}
\resizebox{8cm}{!}{
\begin{tabular}{cccccc}
\, & \multicolumn{4}{c}{\bf Phantom scalar field}   \\
\, & \, & \, & \, & \,   \\
\hline
\, & \, & \, & \, & \,    \\
$Id$ & $\langle x \rangle$ & $\tilde{x}$ & $68\% \ {\rm CL}$  & $95\% \ {\rm CL}$  \\
\hline \hline
\, & \, & \, & \, & \,  \\
\hline \, & \multicolumn{4}{c}{SNIa /GRBs/H(z)}  
 \\
\hline
\, & \, & \, & \, & \,  \\
$H_0$ &0.95 &0.96& (0.93, 1.2) & (0.90, 1.3)  \\
\, & \, & \, & \, & \, \\
$\alpha$ &7.8& 7.9& (5.8,  9.3) & (4.2, 10.3) \\
\, & \, & \, & \, & \,   \\

$h$ &0.72& 0.72 & (0.65, 0.78) & (0.63, 0.80)   \\
\, & \, & \, & \, & \,  \\
\hline
\end{tabular}}
\end{center}
\caption{Constraints on the phantom scalar field parameters from different data samples (combined SNIa  and  GRB Hubble diagrams,  and $H(z)$ data sets. Columns show the mean $\langle x \rangle$ and median $\tilde{x}$ values  and the $68\%$ and $95\%$
confidence limits.}  \label{tab2}
\end{table*}

\section{Discussion and Conclusions}
\label{conclusion}
We analyzed  non-flat cosmological models with an interacting 
quintessence component, where, in turn,  a standard  or a phantom scalar field interacts with the dark matter term. These models are usually characterized by a phenomenological choice of the form of the interaction. Instead, we used the Noether symmetry approach to select the analytical form of both the scalar-field self-interaction potential and the interaction term. It turns out that this latter cancels out  the contributions due to the spatial curvature and reduces the dynamic effect of  cosmological fluid to that of  dust. Of course, there are still effects of  curvature in the evolution of the scalar field, as it can be inferred from the definition of the scalar field effective density and pressure. 

Furthermore, we were able to obtain exact solutions of the cosmological equations, some of which can reproduce the accelerated expansion of  the Universe, both in the case of a standard and a phantom scalar field. Moreover, some solutions make evident the so called super-quintessential behaviour of the equation of state (i.e: $w^{eff}_{\phi} <-1$) due to the coupling term. Finally, we  showed that some of the exact solutions are compatible with different observational dataset related to the background expansion,  as the SNeIa Pantheon data, a GRBs Hubble diagram, and  direct measurements of the Hubble parameter. 
In a forthcoming paper, we are going to perform a  detailed analysis of the  interacting dark energy on the  large scale structures. This approach can allow us to achieve a reliable cosmic history at different redshifts.    
\section*{Acknowledgements}
EP and SC acknowledge the Istituto Nazionale di Fisica Nucleare,  Sez. di Napoli, ({\it Iniziativa Specifica} QGSKY) for the support.  SC acknowledge  also the  Gruppo Nazionale di Fisica Matematica (Istituto Nazionale di Alta Matematica).

\end{document}